\documentclass[%
 reprint,
showpacs,
 amsmath,
 amssymb,
 aps,
pre,
nobalancelastpage]{revtex4-1}

\usepackage{color}
\usepackage{graphicx}
\usepackage{dcolumn}
\usepackage{bm}
\usepackage{esint}
\usepackage{todo}
\usepackage[colorlinks,allcolors=black]{hyperref}
\usepackage[capitalize]{cleveref}
\usepackage{csquotes}
\usepackage{verbatim}
\usepackage{hhline}
\usepackage[normalem]{ulem}
\usepackage[english]{babel} 
\usepackage{microtype}
\usepackage{siunitx}
\usepackage{dsfont}
\usepackage[dvipsnames]{xcolor}
\usepackage{eucal}

{}

\crefname{appsec}{appendix}{appendices}

\begin{document}

\title{Equilibrium return times of small fluctuating clusters and vacancies}

\author{Francesco Boccardo}
\author{Younes Benamara}
\author{Olivier Pierre-Louis}
\email{olivier.pierre-louis@univ-lyon1.fr}
\affiliation{Institut Lumi\`ere Mati\`ere, UMR5306 Universit\'e Lyon 1 - CNRS, 69622 Villeurbanne, France}
\date{\today}
\begin{abstract}
{
The expected return time of a fluctuating two-dimensional cluster or vacancy to a given configuration  
is studied in thermodynamic equilibrium.
We define a family of bond-breaking models
that preserve the number of particles.
This family includes edge diffusion and surface diffusion inside vacancies
in the limit of fast particle diffusion and slow attachment-detachment kinetics.
Within the frame of these bond-breaking models,
the expected return time is found to depend on the 
energies of the configurations and on 
the energies of the excited states formed by removing 
a single particle from the cluster.
High and low temperature regimes are studied.
We clarify the conditions under which 
the return time is a non-monotonous function of temperature:
a minimum is found when
the energy obtained by the average over the excited states of the configuration
weighted by their attachment probabilities
is lower than the energy averaged over all states.
In addition, we show that the optimal
temperature at which the return time is minimum
is shifted to a higher temperature as compared to the 
temperature at which the equilibrium probability is maximum.
This shift is influenced by the average curvature of the cluster
edge, and is therefore larger for vacancies.
}
\end{abstract}

\maketitle



\section{ Introduction}

Two-dimensional monolayer clusters of atoms or particles have been studied 
extensively in the past decades. Since the 1990s, 
advances in visualization techniques such as Scanning Tunnelling
Microscopy for atomic monolayer clusters~\cite{Giesen2001,Jeong1999} and confocal microscopy
for colloid monolayer clusters~\cite{Ganapathy2010}  have enabled their accurate
observation up to the the atom or particle scale.
These observations led to the characterization
of equilibrium shape fluctuations~\cite{Giesen2001,Jeong1999,Pai1997,Zaum2011}
that are caused by the random diffusion of 
their constituent atoms or particles,
both for monolayer clusters~\cite{Pai1997,Zaum2011,Wen1996},
and monolayer vacancies~\cite{Esser1998,Lake2008,Morgenstern2001,Morgenstern1995,Schloer2000,Watanabe1997,Pai2001,Wen1996}.

The static equilibrium properties of these fluctuations
obey well-known equilibrium statistical mechanics~\cite{Saito1996},
and their experimental observation can be used to determine energetic properties
of the cluster edge in atomic 
clusters~\cite{Schloesser1999,Kodambaka2003} and colloid cluster~\cite{Hilou2018,Ou2020}.
In contrast, the dynamical properties of the fluctuations are sensitive to the 
the kinetics of the relevant mass transport mechanisms.
Theoretical investigations of the dynamics of fluctuations
have been developed using 
Langevin models~\cite{Khare1995,PierreLouis2001,Misbah2010}
or Kinetic Monte Carlo simulations
of lattice models  for clusters~\cite{Altaf2006,Wen1996,Lai2017,Bogicevic1998,Mills1999,Combe2000,Uche2009,Hubartt2015}
and vacancies~\cite{Shen2007,Basham2004,Morgenstern2002}.
These modeling studies mostly aimed at predicting the diffusion of the whole cluster
or the time correlation functions of edge fluctuations.
In the following, we focus on a different 
kinetic property of small fluctuating clusters: their time of return to a given configuration.
This focus is motivated by our recent
attempt to describe first passage times 
from one arbitrary configuration to another~\cite{Boccardo2022}.
These first passage times were
evaluated numerically using iterative evaluation 
techniques in Ref.~\cite{Boccardo2022}. 
The expected return time from a configuration to itself 
exhibits similar properties as first passage times~\cite{Boccardo2022} 
and is an easier starting point for numerical and analytical
approaches.

In the present paper, we therefore report on the evaluation of the equilibrium
expected return time to a given configuration.
We work with an extended 
class of models that allows us to 
discuss not only case of clusters with edge diffusion~\cite{Boccardo2022}, 
but also the case of particle diffusion inside vacancies with slow attachment-detachment kinetics.
The expressions derived here lead to simpler and faster estimates
of the expected return times as compared
to the numerical evaluation based on 
iterative evaluation~\cite{Boccardo2022}. They also
lead to an easier and generalized analysis of the high temperature
expansion discussed in Ref.~\cite{Boccardo2022}.
Furthermore, they allow us to perform low-temperature expansions
to determine activation energies. Finally, 
the analysis of the expression of the return time 
provides a simple interpretation for the 
appearance of an optimal temperature at which the 
return time is minimum for configurations with a low energy.

We start in \cref{s:Kac_and_equil} by showing that
the expected return time depends on the expected residence
time in the configuration and on the equilibrium distribution via the well-known Kac lemma~\cite{Kac1947,Aldous2014}.
We distinguish two different definitions of the return time
that differ only by the fact of taking into account the 
time spent on the configuration itself. 

In section \cref{s:broken_bond_models}, we define a class of broken-bond dynamical models with
single particle moves that preserve the cluster area.
We focus on two specific models belonging to this class,
namely clusters with Edge Diffusion (ED) 
and vacancies with Detachment, Diffusion inside the vacancy, and reAttachment (DDA) of atoms or particles
in the limit of slow interface kinetics. 

A discussion of the equilibrium distribution
is then reported in \cref{s:equil_distrib}.
High and low temperature expansions are presented.
We notice that configurations with a low energy
exhibit an optimal temperature for which their equilibrium probability
is maximum. This optimal equilibrium temperature corresponds to the temperature
at which the energy of the configuration is equal to the 
thermodynamic average of the energy. The maximum exists 
only when the energy of the configuration is 
lower than the average of the energy over all configurations.

The return time to a given configuration is discussed in \cref{s:return_times}.
Within the broken-bond models, the expression of the residence
times can be written as a function of the energies
of the excited states that are obtained by 
removing a single atom from the cluster.
High and low temperature expansions of the return times are
discussed.  
We clarify the physical origin of the 
optimal temperature at which large enough low-energy clusters
were found to exhibit a minimum return time in Ref.~\cite{Boccardo2022}.
We show that
a minimum is present when the average energy of the excited states of the configuration
weighted by attachment probabilities
is lower than the average energy over all configurations.
This return time optimal temperature is found to be shifted to higher temperatures as compared
to the optimal temperature of the equilibrium distribution.
In addition, we show that the change of sign of the 
edge curvature between clusters and vacancies
leads to an increased shift towards high temperatures for vacancies.
Finally, the two different definitions of the return time are seen to be similar in
most cases, except for very small clusters and for 
square islands of arbitrary size at low temperatures.

\section{ Relation between 
mean return time, residence time, and equilibrium distribution}
\label{s:Kac_and_equil}

Let us start with some general
and standard relations to define our
cluster as a dynamical stochastic system
at equilibrium.

In the following, the configuration (or shape) of the cluster
is called the state of the cluster, and denoted by $s$.
The physical behavior of the system is described by the rates $\gamma(s,s')$ 
for the transition from a state $s$ of the cluster 
to a state $s'$. We assume Markovian dynamics, and the 
system obeys the master equation
\begin{align}
    \partial_tP(s,t)=\sum_{s'\in {\cal B}_s}\left[
    \gamma(s',s)P(s',t)-\gamma(s,s')P(s,t)
    \right]\, ,
    \label{eq:Master_Equation}
\end{align}
where $P(s,t)$ is the probability that the system 
is in state $s$ at time $t$, and 
${\cal B}_s$ is the set of all states different from $s$ that can be reached
in one transition from $s$.

In equilibrium, the probability $P_{eq}(s)$
to be in state $s$ is independent of time and obeys
\begin{align}
    P_{eq}(s)=\frac{{\rm e}^{-H_s/T}}{\displaystyle \sum_{s'\in{\cal S}}{\rm e}^{-H_{s'}/T}}\, ,
    \label{eq:P_eq}
\end{align}
where $H_s$ is the Hamiltonian, i.e., the energy of the state $s$,
 $T$ is the temperature in units where the Boltzmann constant is equal to $1$,
 and ${\cal S}$ is the set of all states, which is assumed to be finite.

We assume that the cluster dynamics is constrained by 
detailed balance
\begin{align}
    P_{eq}(s)\gamma(s,s')=P_{eq}(s')\gamma(s',s)\, ,
    \label{eq:detailed_balance}
\end{align}
which directly enforce stationarity in 
\cref{eq:Master_Equation}. In addition,
we have from the combination of \cref{eq:P_eq,eq:detailed_balance}
\begin{align}
    \gamma(s,s'){\rm e}^{-H_s/T}=\gamma(s',s){\rm e}^{-H_{s'}/T}.
    \label{eq:detailed_balance_rates_energies}
\end{align}

Let us now consider a situation where we observe the system
during a long time, paying particular attention 
to a given state $s$. Due to ergodicity,
the state $s$ will be reached. After reaching the state $s$, 
the cluster will visit other states
different from $s$. Later,
the system will come back again to $s$, and so on.
Let us denote $t^{\mathrm{obs}}$ the total observation time
and $t^{\mathrm{tot}}(s)$ the total time spend on state $s$
during this observation time. It is clear that since
we are in equilibrium, the equilibrium probability $P_{eq}(s)$
is the fraction of time spent in the state $s$, 
and is therefore 
\begin{align}
    P_{eq}(s)=\lim_{t^{\mathrm{obs}}\rightarrow\infty}\frac{t^{\mathrm{tot}}(s)}{t^{\mathrm{obs}}}.
    \label{eq:eq_obs_residence}
\end{align}
Then, let us define the loop number $n_{\mathrm{\ell}}$ as the number of times
the system passes on the state $s$. 
The expected residence time $t(s)$ in state $s$ obeys~\footnote{
Here and in the following, {\it expected} refers to an expected value and is a synonym of {\it average}.}
\begin{align}
    t(s)=\lim_{t^{\mathrm{obs}}\rightarrow\infty}\frac{t^{\mathrm{tot}}(s)}{\displaystyle n_{\mathrm{\ell}}}.
    \label{eq:def_t_0}
\end{align}
Remark that the expected residence time  of a state $s$ 
is the average of the time before a transition to a different state $s'\neq s$.
As a consequence, the transitions that directly take the system
from $s$ to itself are discarded and are not considered as loops.

Moreover, we define the expected loop time $\tau^{\mathrm{\ell}}(s)$
as the average period of the return to the state $s$, so that
\begin{align}
    \tau^{\mathrm{\ell}}(s)
    =\lim_{t^{\mathrm{obs}}\rightarrow\infty}\frac{t^{\mathrm{obs}}}{n_{\mathrm{\ell}}}.
    \label{eq:def_tau_loop}
\end{align}
Combining \cref{eq:eq_obs_residence,eq:def_t_0,eq:def_tau_loop}, we obtain
a form of the well known Kac lemma~\cite{Kac1947,Aldous2014}
\begin{align}
    \tau^{\mathrm{\ell}}(s)=\frac{t(s)}{P_{eq}(s)}\, ,
    \label{eq:loop_res_eq}
\end{align}
that relates the loop time $\tau^{\mathrm{\ell}}(s)$
to the stationary distribution $P_{eq}(s)$.
In the literature, $\tau^{\mathrm{\ell}}(s)$ is often called the 
expected return time to state $s$.
However, following the definition of Ref.~\cite{Boccardo2022}, 
we define the expected return time $\tau^{\mathrm r}(s)$
as the expected time spent outside state $s$ before returning to it.
We therefore have
\begin{align}
   \tau^{\mathrm r}(s)=\tau^{\mathrm{\ell}}(s)-t(s)=t(s)\left(\frac{1}{P_{eq}(s)}-1\right).
    \label{eq:return_res_eq}
\end{align}
In \cref{eq:loop_res_eq,eq:return_res_eq}, the equilibrium distribution
is given by \cref{eq:P_eq}, and the residence time can be written directly
as a function of the rates
\begin{align}
    t(s)=\frac{1}{\displaystyle \sum_{s'\in {\cal B}_s}\gamma(s,s')}.
    \label{eq:t0_gamma}
\end{align}
Note that our definitions
imply that physical events that do not change the state
$s$ of the system are not listed in the transitions in \cref{eq:t0_gamma}.

As a summary, the knowledge of the energies $H_s$ and of the rates $\gamma(s,s')$ allows one to 
determine the expected return time.
In order to analyze the consequences
of this simple result for thermal fluctuations 
of few-particles clusters, we need to define
more precisely the kinetics of the model,
which is determined by the rates $\gamma(s,s')$.

\section{Broken-bond models}
\label{s:broken_bond_models}

\subsection{Model definitions}

\begin{figure*}
    \centering
    \includegraphics[width=.75\linewidth]{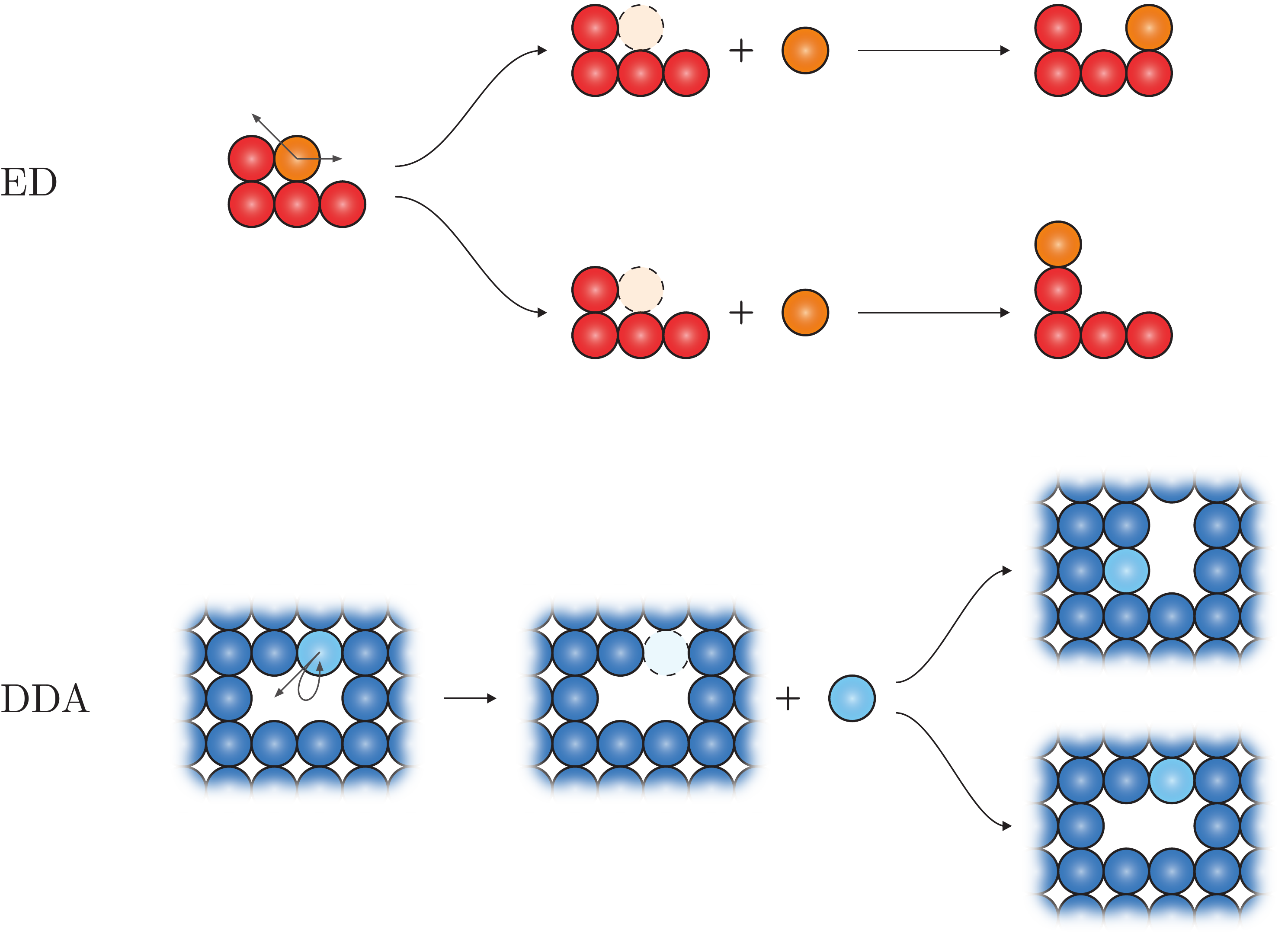}
\caption{ Schematics of particle moves within the two broken-bond models discussed here.
In red, cluster with Edge Diffusion (ED).
In blue, vacancy with particle Detachment-Diffusion-reAttachment (DDA). 
In each case, the arrows in the leftmost schematic indicate the possible moves
of the atom in lighter color.
Intermediate states after the detachment
of one particle are called excited states.
The states after one move are shown in the rightmost panels.
Note that the final state in the lower right panel is identical
to the initial state.
}
    \label{fig:schematics}
\end{figure*}

We consider a two-dimensional cluster on a square lattice with lattice parameter $a$
and nearest-neighbor bond energy $J$. 
A cluster  is defined as a part of the lattice 
composed of $N$ sites that are connected to each other through nearest-neighbors.
Two cluster configurations are considered to be different if they cannot be obtained 
one from the other via translations.
This makes our definition of cluster states identical to that
of free polyominoes or free lattice animals~\cite{Guttmann2009}.
Vacancies are defined in a similar way as 
a set of empty connected sites in a full monolayer.

The dynamics is assumed to result from moves that involve
only one particle at a time.
The number of different moves that can take the system from state $s$
to a state $s'$ is denoted as $k_{ss'}$.
These moves from $s$
to $s'$ are indexed by $k=1,..,k_{ss'}$.
We assume that each move has a reverse move. Hence,
the reverse moves can also be indexed by the same index $k$,
and  the total number of moves between the states $s$ and $s'$
is the same than the number of reverse moves
\begin{align}
k_{ss'}=k_{s's}.
\end{align}
Following usual models for activation
of particle or atom diffusion~\cite{Gilmer1972a,Gilmer1972b,Kotrla1996},
the rate of the transition from state $s$ to state $s'$ due to the $k$-th 
one-particle move 
is assumed to take an Arrhenius form
\begin{align}\label{eq:TST_hopping}
    \gamma_k(s,s') = \nu \,b_{ss';k}\,{\rm e}^{-n_{ss';k}J/T},
\end{align}  
where $\nu$ is an attempt frequency,  
$n_{ss';k}$ is the number of in-plane nearest neighbors 
of the moving particle in state $s$ before hopping, and $b_{ss';k}$
is a model dependent attachment probability.
The total transition rate from $s$ to $s'$ therefore reads
\begin{align}\label{eq:TST_hopping_total}
    \gamma(s,s') =\sum_{k=1}^{k_{ss'}}\gamma_k(s,s') .
\end{align}  
In addition, for consistency, the rates vanish
for moves that are not authorized, i.e., $\gamma(s,s')=0$ when $k_{ss'}=0$.

The rates \cref{eq:TST_hopping} define 
a family of models which correspond to 
different ways of setting the possible moves and the associated parameters $b_{ss';k}$.
Each physical model for the re-attachment of the
particles after detachment provides a specific expression of $b_{ss';k}$.
Here we consider four basic constraints on the 
reattachment rules.
First, particles always re-attach 
after detachment, so that the number of 
particles in the cluster is not changed.
Second, particles reattach instantaneously.
This means that we assume that the time
needed for the reattachment process is negligible
as compared to the time of detachment.
Third, we assume that all detachment-reattachment
moves have a reverse move.
This condition is necessary
to enforce detailed balance at equilibrium \cref{eq:detailed_balance}.
Finally, the detachment-reattachment moves do not
break the cluster into disconnected clusters.

Within this family of detachment re-attachment processes, we wish to focus
on two types of dynamics described schematically in \cref{fig:schematics}. The first 
model is Edge-Diffusion (ED) dynamics of particle clusters.
Various models have been
developed in order to describe ED in
atomic monolayer clusters (see, e.g., \cite{Lai2017,Sanchez1999} and references
therein).
Here, we choose to use a simple model following the rules
of Refs.~\cite{Boccardo2022,PierreLouis2000}. In this case, the particles
reattach to nearest neighbor or next-nearest neighbor sites
along the edge of the cluster, i.e. to sites that have at least one
nearest-neighbor bond with another particle of the cluster.
Moreover, we choose
\begin{align}
    b_{ss';k}=1.
\label{eq:b_ssk_ED}
\end{align} 
With this choice, we aim at describing 
the different possible moves of a particle as independent
processes.

The second model accounts for Detachment, Diffusion, and reAttachment (DDA)
of particles inside vacancies. In this case, the moves
are composed of a chain of three processes: first detachment, then diffusion,
and then reattachment. Reattachment can occur
to any site at the edge inside the cluster with the same probability. 
This model corresponds to a regime where re-attachment is slower than 
diffusion, so that particles have ample time to
diffuse inside the cluster before re-attaching.
In such a situation, the probability of presence of the atom
inside the cluster has time to relax to a spatially homogeneous
probability, and as a consequence, we can choose an attachment probability
that is independent of the position of the attachment site~\footnote{
Such a model can for example be obtained in the limit
of lattice models with a small attachment rate,
which correspond to the limit of small $Q$ in Refs.~\cite{Gagliardi2022,Reis2022}. }.
We therefore have
\begin{align}
    b_{ss';k}=\frac{1}{\ell_{ss';k}^\dagger}\, , 
\label{eq:b_ssk_DDA}
\end{align}
where $\ell_{ss';k}^\dagger$ is the number
of sites for the particle to re-attach along the edge
of the cluster. As an example, for the move of the dimer vacancy on \cref{fig:schematics},
there are two possible attachment sites
(described by the red arrows) so that $\ell_{ss';k}^\dagger=2$.

For consistency, we also forbid configurations
where an isolated, non-mobile particle is inside the vacancies. 
This condition adds novel constraints on the moves
that are imposed in addition to the non-breaking constraints.
It also leads to a reduced number of states 
for vacancies as compared to clusters.

\begin{figure}
    \centering
    \includegraphics[width=\linewidth]{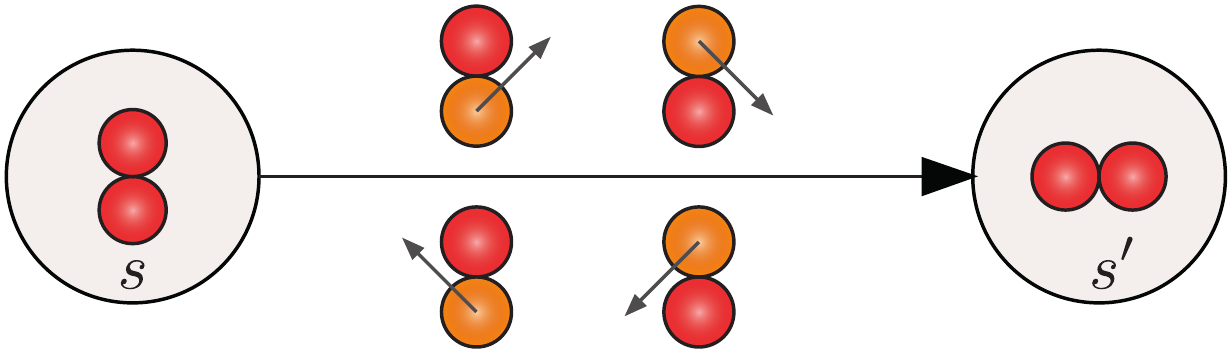}
    \caption{Different particle moves can lead to the same
    transition. Here, we consider a dimer cluster $N=2$ with edge diffusion.
    The transition from one state
    to the other can be achieved in 4 different ways. Hence $k_{ss'}=4$.}
    \label{fig:k_ss_dimer}
\end{figure}

\begin{figure}
    \centering
    \includegraphics[width=\linewidth]{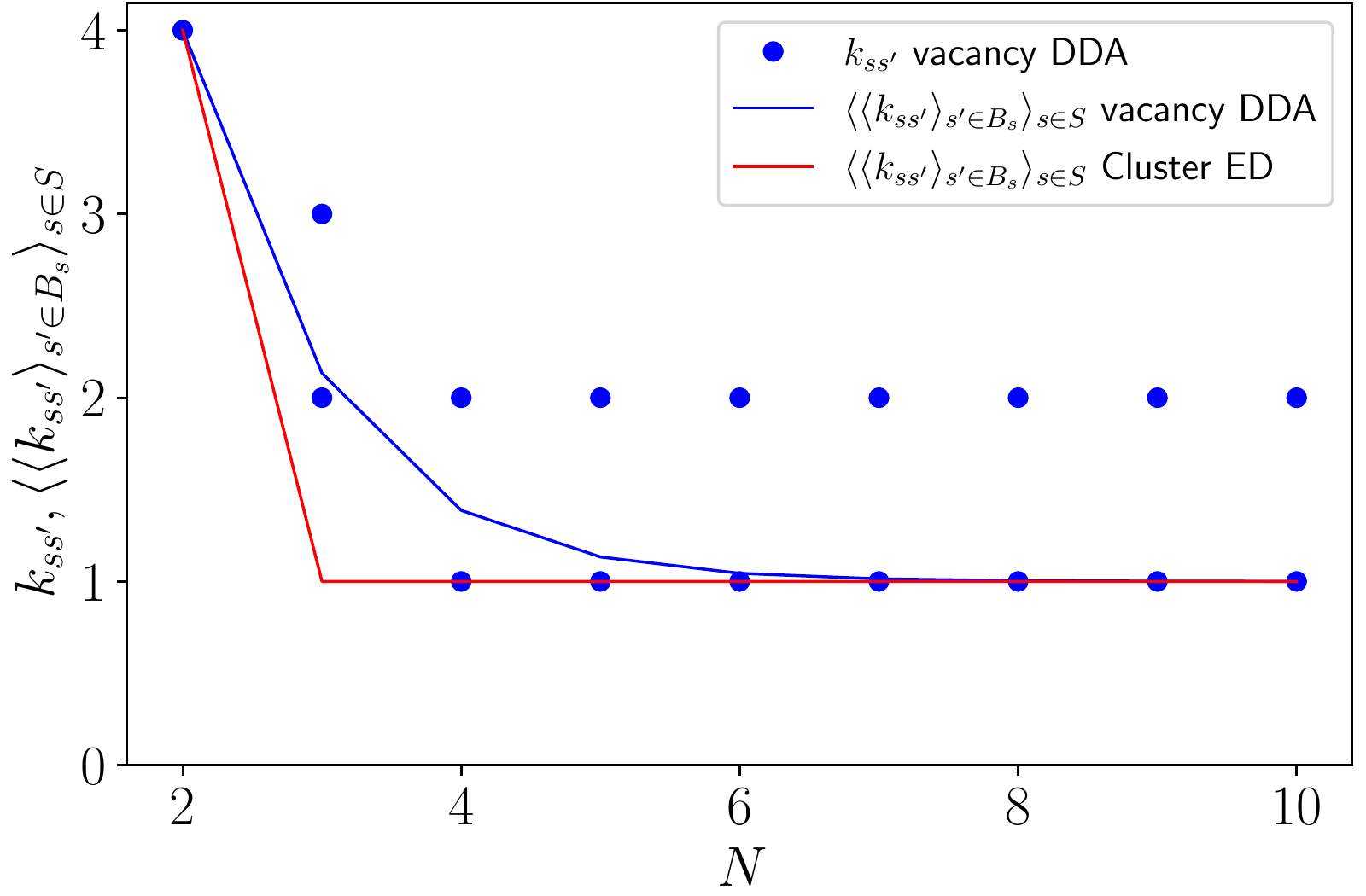}
    \caption{Number of different moves $k_{ss'}$ to go from a state $s$ to
    another state $s'$ as a function of the cluster size $N$. The symbols show all possible values of $k_{ss'}$ 
    within the two models cluster ED and vacancy DDA.
    The solid lines correspond to the average value $\langle\langle k_{ss'}\rangle_{s'\in{\cal B}_s}\rangle_{s\in\cal S}$
    of $k_{ss'}$ over all possible moves.}
    \label{fig:k_ss}
\end{figure}

The total number of moves $k_{ss'}$ 
from state $s$ to state $s'$ depends on the model and can be larger than $1$.
For example, there are 4 possible moves for the transition 
of a dimer cluster with ED shown in \cref{fig:k_ss_dimer}.
The dependence of $k_{ss'}$ on $N$ is summarized in \cref{fig:k_ss}.
In the cluster ED model, $k_{ss'}=4$ for 
the two transitions of the dimers
with $N=2$ particles, and $k_{ss'}=1$ for $N\geq 3$, as already noticed in Ref.\cite{Boccardo2022}.
In the vacancy DDA model, $k_{ss'}$
is decreasing as $N$ increases, but there are 
moves with $k_{ss'}=2$ for any $N$.
However, as seen in \cref{fig:k_ss}, these moves are rare
in the sense that
the average $\langle\langle k_{ss'}\rangle_{s'\in{\cal B}_s}\rangle_{s\in{\cal S}}$ over all possible moves
tends quickly to $1$ as the number $N$ of particles
in the cluster increases.
For any set of states ${\cal Z}$, we have defined the average of a state-dependent function $f_s$ as
\begin{align}
    \langle f_s \rangle_{s\in {\cal Z}}=\frac{1}{|{\cal Z}|}\sum_{s\in {\cal Z}}f_s
\end{align}
with $|{\cal Z}|$ the cardinal of ${\cal Z}$.
Moreover, we recall that ${\cal S}$ is the set of all possible states for 
a fixed size $N$,
and ${\cal B}_s$ is the set of all states different from $s$
that can be reached in one transition from $s$.

\subsection{Hamiltonian and excited states}

The broken bond model is directly related
to the Ising or lattice-gas lattice models.
As discussed e.g. in Refs.~\cite{Saito1996,Livi2017},
the energy of a cluster in a state $s$ is simply related to the length $L_s$
of its edge 
\begin{align}
H_s=\frac{J}{2}\,\frac{L_s}{a}\, .
\label{eq:Hs_Ls}
\end{align}
A simple intuitive interpretation of this result
is that the breaking of a bond costs an energy $J$ and
increases the length of the edge by $2a$.
The energy cost for the formation
of an elementary segment of the edge of length $a$
is therefore $J/2$, and 
we recover \cref{eq:Hs_Ls}.

As seen from \cref{fig:schematics}, an atomic move which allows for the transition from state $s$ 
to state $s'$ can be decomposed into two stages. 
In the first stage, we create an excited state by breaking all the bonds
that the atom has in its initial condition. This excited state can be interpreted as the situation where the particle 
is brought to the saddle point of the diffusion energy landscape.
In the second stage, we reattach the atom.

Let us denote excited states with a $\dagger$ symbol.
We define the Hamiltonian of the excited state 
$H_{ss';k}^\dagger$ as the 
Hamiltonian of the state obtained by removing the detaching particle
during the $k$-th particle move 
that leads to the transition from $s$ to $s'$. 
We then have
\begin{align}
H_{ss';k}^\dagger=\frac{J}{2}\,\frac{L_{ss';k}^\dagger}{a}+2J
\label{eq:Hamiltonian_dagger}
\end{align}
where $L_{ss';k}^\dagger$ is the length of the edge of the cluster
obtained by removing the moving atom. In the 
examples shown in \cref{fig:schematics},
$L_{ss';k}^\dagger=10$ for cluster ED, and 
$L_{ss';k}^\dagger=8$ for vacancy DDA.
The additional constant $2J$ in \cref{eq:Hamiltonian_dagger}
accounts for the energy of the detached particle. Indeed, since
a detached particle has $4$ broken bonds and each
broken bond costs an energy $J/2$, its total
broken-bond energy is $4J/2=2J$.
Note that the excited state for the $k$-th transition $s\rightarrow s'$ is the same
as the excited state for the $k$-th transition $s'\rightarrow s$. 
We therefore
have
\begin{align}
L_{ss';k}^\dagger=L_{s's;k}^\dagger \, ,
\quad
H_{ss';k}^\dagger=H_{s's;k}^\dagger \,.
\label{aeq:symm_H_dagger}
\end{align}

The change of 
cluster edge length when removing a single atom is related to the 
number bonds $n_{ss';k}$ that are broken via~\cite{Saito1996}
\begin{align}
L_{ss';k}^\dagger-L_s=2a(n_{ss';k}-2)\, .
\end{align}
As a consequence, the activation energy of \cref{eq:TST_hopping} reads
\begin{align}
n_{ss';k}J=2J+\frac{J}{2a}(L_{ss';k}^\dagger-L_s)=H_{ss';k}^\dagger-H_s\, .
\label{eq:n_H}
\end{align}
This latter equation 
indicates that the product $n_{ss';k}J$
can be decomposed into two parts: 
the excited state energy $H_{ss';k}^\dagger$ which is
identical for the move and its reverse, 
and the energy of the initial state $H_s$.
In other words, the energy barrier for detachment
is the difference between the energy of the excited state,
which play the role of an effective saddle point for the energy,
and the energy of the initial state. Combining \cref{eq:TST_hopping,eq:TST_hopping_total,eq:n_H}
the rates may now be written as
\begin{align}\label{eq:rates_Hdag_Hs}
    \gamma(s,s') = \nu \, {\rm e}^{H_s/T}\sum_{k=1}^{k_{ss'}} 
    b_{ss';k} \,{\rm e}^{-H_{ss';k}^\dagger/T}.
\end{align}  

Moreover, 
the cluster ED and vacancy DDA models both obey the 
symmetry relation
\begin{align}
    b_{ss';k}=b_{s's;k}\, .
\label{eq:b_ssk_symm}
\end{align} 
This relation is trivially valid for cluster ED as seen from \cref{eq:b_ssk_ED}.
In the case of vacancy DDA, the number of sites $\ell_{ss';k}^\dagger$ to which detached 
particles can be re-attached is a property
of the $k$-th excited state between the states 
$s$ and $s'$. As a consequence, we have $\ell_{ss';k}^\dagger=\ell_{s's;k}^\dagger$
in \cref{eq:b_ssk_DDA}
and the symmetry property \cref{eq:b_ssk_symm} follows.

Since \cref{eq:rates_Hdag_Hs} is the product 
of ${\rm e}^{H_s/T}$ with a factor that is 
invariant under the exchange of indices $s\leftrightarrow s'$,
the rates $\gamma(s,s')$ are seen to obey detailed balance \cref{eq:detailed_balance_rates_energies}.
Hence, the broken-bond models exhibit a well defined 
equilibrium state characterized by the energies \cref{eq:Hs_Ls}.

In the following, 
we will use units where $J=1$, $a=1$ and $\nu=1$.

\section{Equilibrium distribution}
\label{s:equil_distrib}

\subsection{Energy levels}

The energy of a cluster can only explore a finite number
of energy levels indexed by $i=0,1,...,i_{\mathrm{max}}$, with energy $H_{(i)}$  
corresponding to a given edge length $L_{(i)}=2H_{(i)}$.
The energy levels obey 
\begin{align}
    L_{(i)}=L_{(0)}+2i, 
    \quad
    H_{(i)}=\frac{L_{(i)}}{2}=H_{(0)}+i,
    \label{eq:energy_levels}
\end{align}
where the ground state energy $H_{(0)}=L_{(0)}/2$ depends on 
the size $N$ of the cluster.

\begin{figure}
    \centering
    \includegraphics[width=\linewidth]{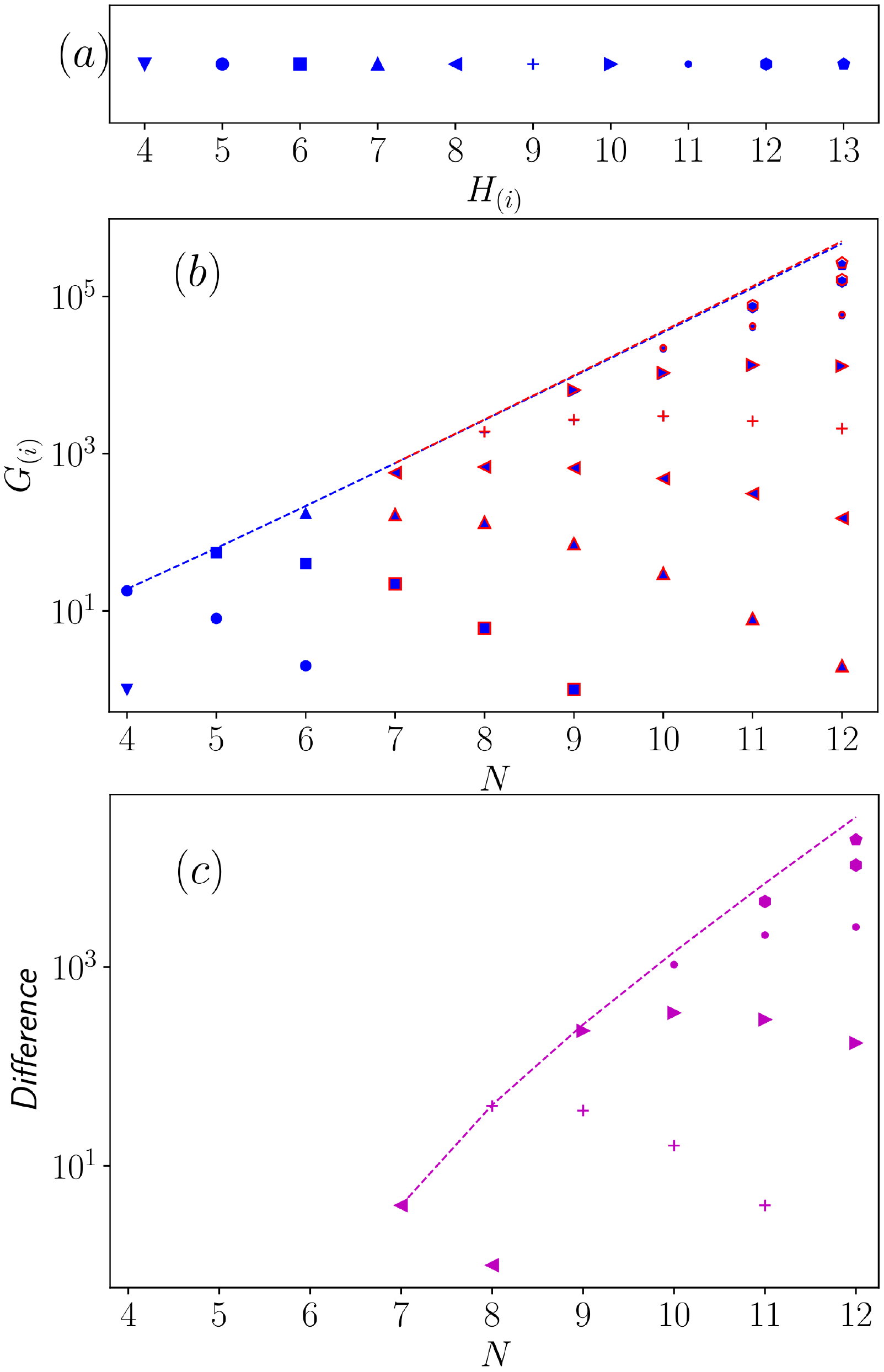}
    \caption{Degeneracy $G_{(i)}$ of different energy levels $i$.
    Each symbol corresponds to a different value of energy, reported
    in the scale in (a).
    (b) Degeneracies $G_{(i)}$. 
    Red empty symbols and blue full symbols respectively correspond to clusters
    and vacancies.
    The total number of states $S_N$ is also reported
    as a dashed line.
    (c) Difference between the degeneracies $G_{(i)}$ of clusters and vacancies.
    }
    \label{fig:degeneracy_Gi}
\end{figure}

The number of states that correspond to
the energy level $i$ is denoted as $G_{(i)}$.
We also define the total number of states 
\begin{align}
S_N=\sum_{i=0}^{i_{\mathrm{max}}}G_{(i)}\, .
\end{align}
For $N=2$ and $N=3$, there is only one energy level and $i_{\mathrm{max}}=0$.
We have  $G_{(0)}=S_2=2$ for $N=2$,
and $G_{(0)}=S_3=6$ for $N=3$.
The values of $G_{(i)}$ and $S_N$ obtained by explicit enumeration for $4 \leq N\leq 12$
are reported in \cref{fig:degeneracy_Gi}.
These numbers are different 
for clusters and vacancies
because 
of the prohibition of isolated particles inside vacancies
which appear for $N\geq 7$.
However, the difference is small as compared to $G_{(i)}$ itself, as seen from \cref{fig:degeneracy_Gi}(b,c).
When $N\gg 1$, we expect $S_N$ to be well approximated by the asymptotic form~\cite{Guttmann2009}
\begin{align}
S_N\approx c\lambda^N/N \, .
\label{eq:S_N_asympt}
\end{align}
The parameters of this asymptotic form are well known for
free polyominoes, which correspond exactly
to our clusters: $ \lambda \approx 4.0626$ and $ c \approx 0.3169$~\cite{Jensen2000}.
Vacancies have a slightly lower $S_N$
the asymptotic form of which is not known.
However, we expect that clusters and vacancies should have
the same value of $\lambda$~\footnote{
Since the number of possible positions of an isolated
particle in a vacancy is at most equal to $N$,
$S_N$ for vacancies is at most $N$
times smaller than $S_N$ for clusters. 
Hence, the elimination of these configurations
can only lead to algebraic (power-law) corrections.
As a consequence,
the exponential growth rate $\lambda$ is the same for
clusters and vacancies.}.

\begin{figure}
    \centering
    \includegraphics[width=\linewidth]{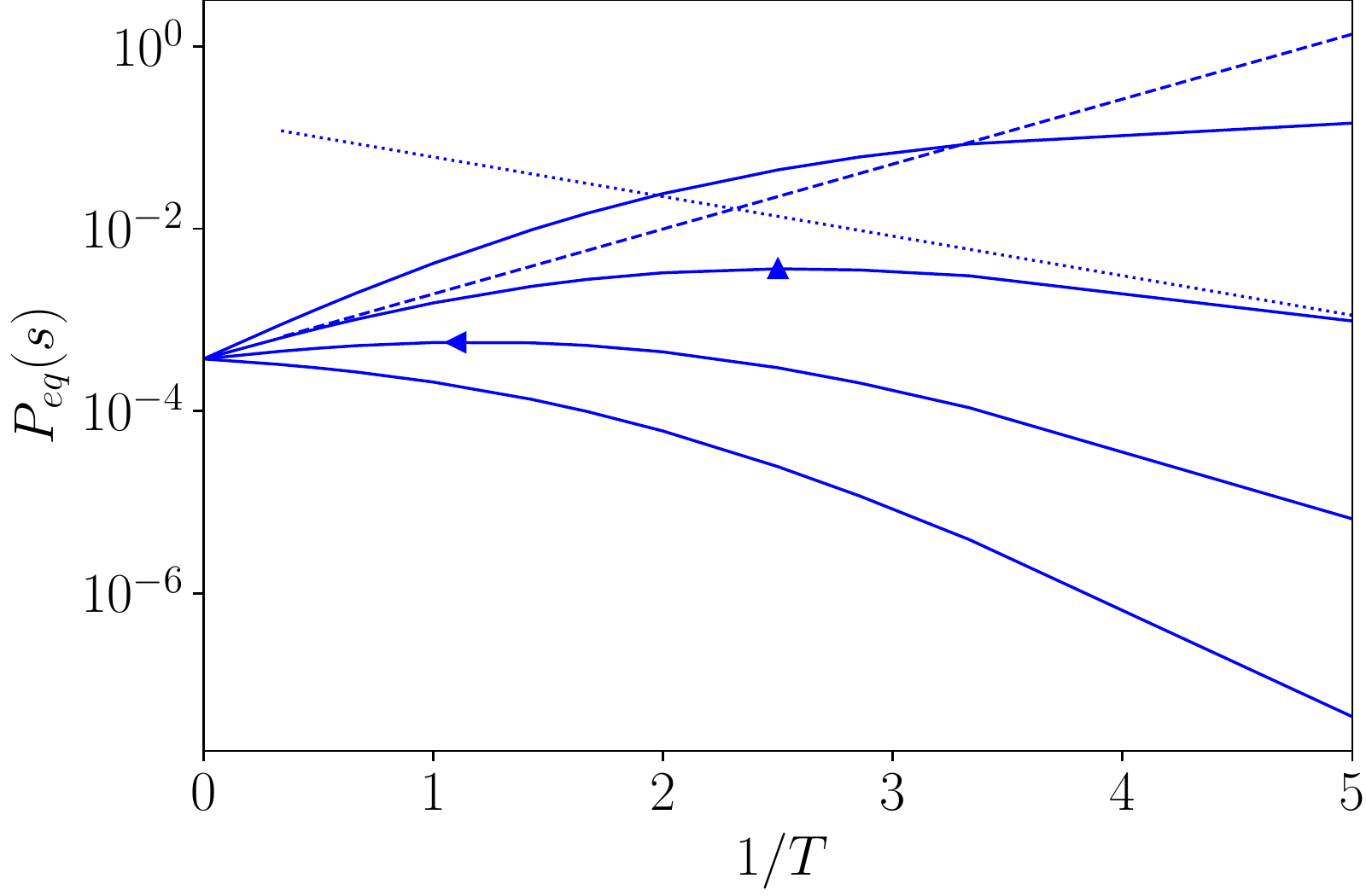}
    \caption{Equilibrium probability distribution for vacancies with $N=8$.
    The solid lines correspond to $P_{eq}$. There are 4 energy levels with $i=0,1,2,3$ from top to bottom,
    corresponding to energies $H_{(i)}=H_{(0)}+i$ where $H_{(0)}=6$ is the ground state energy.
    The dashed and dotted lines respectively report the high and low temperature expansions.
    Symbols indicate the maximum of $P_{eq}$. 
    }
    \label{fig:p_eq}
\end{figure}

In \cref{fig:p_eq}, the equilibrium probability distribution
$P_{eq}(s)$ given by \cref{eq:P_eq} is plotted
as a function of the inverse temperature for various energy levels $i$.
The case reported in \cref{fig:p_eq} corresponds to 
a vacancy octamer $N=8$, with four energy levels $i=0,1,2,3$.
The case of a cluster with $N=8$ is not plotted but is very similar.

As expected,  the equilibrium distribution decreases
as $i$ increases. However, the temperature
dependence of $P_{eq}$ is less trivial. In \cref{fig:p_eq}, $P_{eq}$ is seen to
increase monotonously as the temperature is decreased for the ground state $i=0$,
and  decreases monotonously for large $i$.
Interestingly, $P_{eq}$ 
exhibit a maximum for intermediate values of $i$.
This  non-monotonic behavior can be understood intuitively. 
Indeed, at  very low temperatures the cluster stays in the ground states $i=0$.
As the temperature increases, higher energy levels are populated
and the corresponding values of $P_{eq}$ increases.
In contrast, in the limit of very high temperatures, all
states are populated equally and $P_{eq}\rightarrow P_\infty$ as $1/T\rightarrow 0$, with
\begin{align}
    P_\infty=\frac{1}{S_N}.
\end{align}
When the temperature is decreased from
this high temperature limit, 
the states with a lower energy become more probable
while those with a higher energy become less probable.
Thus, since their probability increases 
when starting both from the very low temperature limit and from
the very high temperature limit, 
low energy states  that are not the ground state must exhibit a maximum at some finite temperature.
We call this temperature the optimal equilibrium temperature $T_{\mathrm{m}}^{eq}$,
because this corresponds to the temperature where the probability of observing
a given state is highest.
The position of the maximum is marked by a symbol in \cref{fig:p_eq}.

\subsection{High and low temperature expansions}

A high temperature expansion of \cref{eq:P_eq} to first order in $1/T$
leads to
\begin{align}
    P_{eq}(s)&\xrightarrow[T\rightarrow \infty]{}
    P_\infty\left[1+\frac{M_{\mathrm{eq}}(s)}{T}\right]\, ,
    \nonumber \\
    M_{\mathrm{eq}}(s)&=\langle H_{s'}\rangle_{s'\in {\cal S}}-H_s\, ,
    \label{eq:Peq_HT}
\end{align}
where the infinite temperature equilibrium distribution $P_\infty={1}/{S_N}$
is independent of the state $s$.

In the opposite limit at low temperatures,
we have
\begin{align}
    P_{eq}(s)\xrightarrow[T\rightarrow 0]{}
    \frac{1}{G_{(0)}}{\rm e}^{(H_{(0)}-H_s)/T}\, ,
    \label{eq:Peq_LT}
\end{align}
where $H_{(0)}$ is the ground state energy of 
the cluster of size $N$ and $G_{(0)}$ is the number of different states that have the ground state
energy $H_{(0)}$.

\subsection{Equilibrium optimal temperature}
\label{s:equil_optimal_temperature}

As seen from \cref{fig:p_eq} and discussed above, the ground state
exhibits a monotonously increasing $P_{eq}$ when decreasing the temperature.
Thus combining \cref{eq:Peq_HT,eq:Peq_LT}, a simple criterion for the presence of a maximum
at finite temperature is that the energy of the state is lower
than the average energy over all states 
\begin{align}
H_{(0)}<H_s<\langle H_s'\rangle_{s\in {\cal S}}.
\label{eq:criterion_max}
\end{align}
This inequality indicates that clusters with a low
energy exhibit a finite temperature $T_m^{eq}(s)$
that maximizes the equilibrium probability $P_{eq}(s)$.
Note that this temperature depends only on the energies of the system, not on the kinetics.

In general, the condition of a maximum of $P_{eq}(s)$ reads $\partial_{T}P_{eq}(s)=0$,
leading to an implicit equation for the optimal equilibrium temperature
\begin{align}
\label{eq:optimal_temp_equil}
    H_s = \overline{H_{eq}}|_{T=T_m^{eq}(s)},
\end{align}
where we have defined the thermodynamic average of the energy
\begin{align}
\overline{H_{eq}}=\sum_{s'\in {\cal S}}H_{s'} P_{eq}(s')\, .
\end{align}

\begin{figure}[h]
    \centering
    \includegraphics[width=\linewidth]{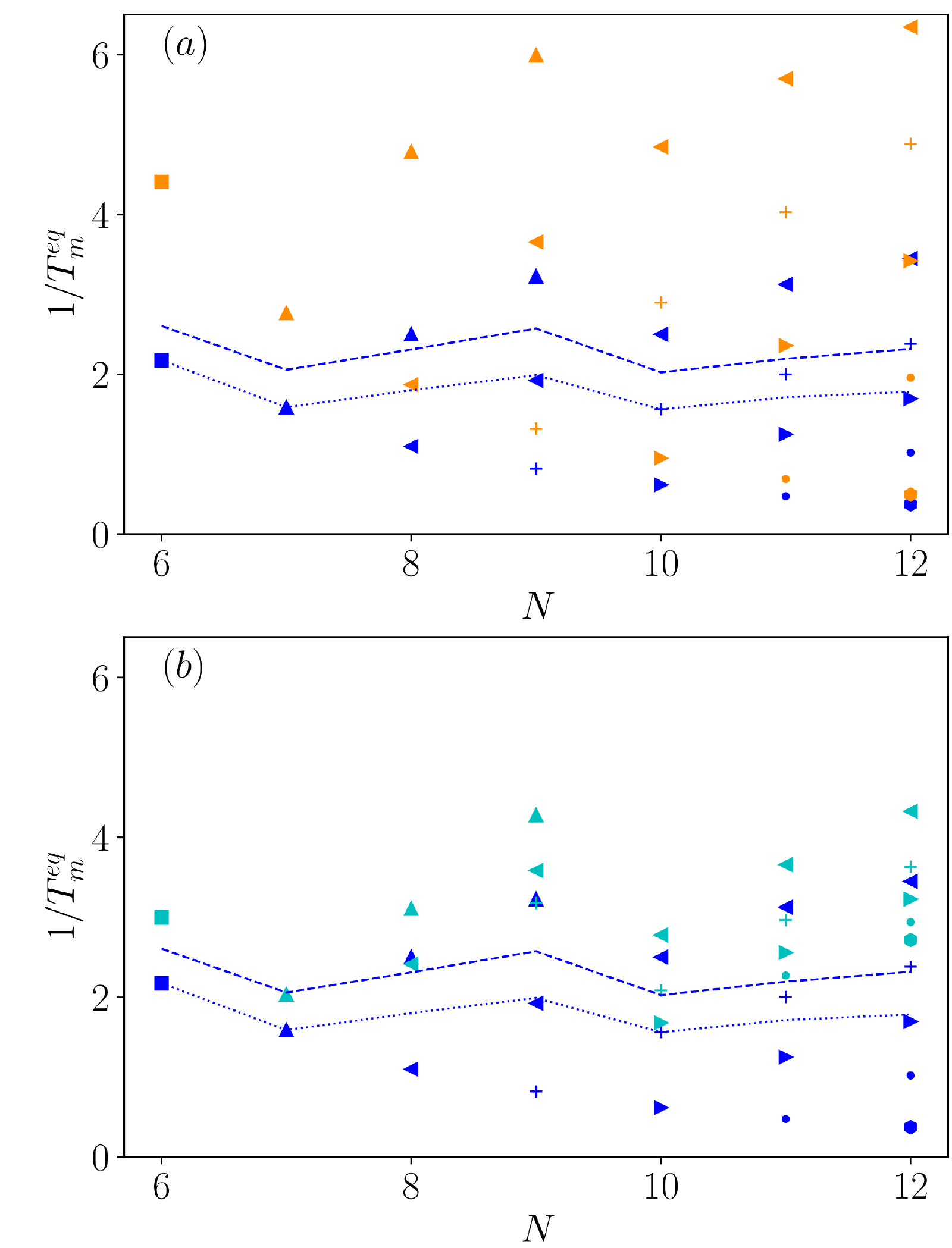}
    \caption{Equilibrium optimal temperature $T_m^{eq}$ as a function
    of $N$ for different energy levels $i$. We show the case of vacancies
    (clusters are very similar).
    The optimal temperature $T_m^{eq}$ are in blue (dark) symbols.
    The symbols correspond to the energies in the scale at the top of \cref{fig:degeneracy_Gi}.
    The dashed lines represent $1/T_a^{eq}$ from \cref{eq:Tm_HT_LT},
    and the dotted line is the average of the inverse optimal temperatures over energy levels.
    In (a), the optimal temperatures are compared to the high temperature
    approximation \cref{eq:Tm_eq_HT} in orange (light) symbols.
    In (b), the optimal temperatures are compared to the low temperature
    approximation \cref{eq:Tm_eq_LT} in cyan (light) symbols.
    }
    \label{fig:Tmax_eq}
\end{figure}

A first estimation of the temperature 
$T_m^{eq}(s)$ at which $P_{eq}(s)$ is maximum
can be obtained from a comparison of 
the high and low temperature expansions \cref{eq:Peq_HT,eq:Peq_LT}.
Assuming that the $T_m^{eq}(s)$ corresponds 
to the temperature $T_{\text{a}}^{eq}(s)$ where both expressions are equal,
and reformulating the high-temperature expansion as 
$P_{eq}(s)\approx S_N^{-1}\exp[M_{\mathrm{eq}}(s)/T]$,
we find
\begin{align}
    T_{\text{a}}^{eq}= 
    \frac{
    \langle H_{s'}\rangle_{s'\in {\cal S}}-H_{(0)}}
    {\displaystyle \ln\left[\frac{S_N}{G_{(0)}}\right]}.
    \label{eq:Tm_HT_LT}
\end{align}
Such a temperature corresponds to the
crossing of the high and low temperature
approximations shown in \cref{fig:p_eq}.
Note that this approximate expression
does not depend on the state $s$, and depend only
on the cluster size $N$.
As seen from \cref{fig:Tmax_eq}, 
the inverse temperature $1/T_{\text{a}}^{eq}$ provides a fair account 
of the average of $1/T_m^{eq}(s)$
over the energy levels for a given cluster size $N$.
However, $T_{\text{a}}^{eq}$ is not an accurate estimate as it does not
account for the strong dispersion of the optimal temperatures
depending on the energy-level $i$.

A high temperature expansion of \cref{eq:optimal_temp_equil}
to first order in the inverse temperature $1/T$ leads to another
approximate expression of the optimal equilibrium temperature
of a state $s$ belonging to the energy level $i$
\begin{align}
    T_{\mathrm{HT}}^{eq}=\frac{\langle H_{s'}^2\rangle_{s'\in{\cal S}}-\langle H_{s'}\rangle^2_{s'\in{\cal S}} } 
    {\langle H_{s'}\rangle_{s'\in{\cal S}}-H_{(0)}-i}\, .
    \label{eq:Tm_eq_HT}
\end{align}
As expected, this estimate is seen
to account well for the optimal temperature when
$1/T_m^{eq}$ is small in \cref{fig:Tmax_eq}(a). However,
this expression underestimates lower optimal temperatures,
which corresponds to  larger values of $1/T_m^{eq}$.

In the opposite limit of low temperatures,
we first notice that the ground state can be considered
as a state with an optimal temperature at zero temperature.
This statement is indeed in agreement with \cref{eq:optimal_temp_equil}
and is intuitively associated to the presence of a horizontal tangent
of the curve $P_{eq}(s)$ in \cref{fig:p_eq}. For states
in higher energy levels $i\geq 1$,
an expansion of \cref{eq:optimal_temp_equil} to first order in the low-temperature small parameter
$\exp[-1/T]$ suggests
\begin{align}
    T_{\mathrm{LT}}^{eq}=\dfrac{1}{\ln \dfrac{G_{(1)}}{iG_{(0)}} }\, .
    \label{eq:Tm_eq_LT}
\end{align}
However, note that the smallness $\exp[-1/T]$
requires that $iG_{(0)}/G_{(1)}\ll 1$.
In general, low temperature expansions are delicate because they require
a precise knowledge of the variation of $G_{(i)}$ with $i$ an $N$,
which is still an open problem~\cite{Guttmann2009}.
The comparison reported in \cref{fig:Tmax_eq}(b) shows
a fair a agreement between \cref{eq:Tm_eq_LT} and the lowest optimal temperatures corresponding to $i=1$. 
However,  $T_{\mathrm{LT}}^{eq}$ is not accurate for higher energy levels.

\section{Return times}
\label{s:return_times}

\subsection{Expression of $\tau^{\mathrm r}(s)$}
\label{s:expression_taur}

Combining \cref{eq:P_eq,eq:return_res_eq,eq:t0_gamma,eq:rates_Hdag_Hs}
we obtain an expression of the expected return time
\begin{align}
    \tau^{\mathrm r}(s)=
    \frac{\displaystyle\sum_{s'\in {\cal S}\backslash s}{\rm e}^{-H_{s'}/T}}
    {\displaystyle\sum_{s'\in {\cal B}_s} \sum_{k=1}^{k_{ss'}}
    b_{ss';k}\,{\rm e}^{-H_{ss';k}^\dagger/T}}\, ,
\label{eq:tau_r_Hs_Hdag}
\end{align}
where ${\cal S}\backslash s$ is the set of all
states but the state $s$.
We recall that our definition of the residence time discards the
moves that take the system from a state $s$ to itself.
An example of a move that does not change the state $s$ 
is shown at the bottom of \cref{fig:schematics}.

In the limit where $T\rightarrow \infty$,
all exponential terms in \cref{eq:tau_r_Hs_Hdag} are equal to $1$. Then \cref{eq:tau_r_Hs_Hdag}
leads to a generalization of the well know
formula~\cite{Lovasz1993} for return times $\tau_\infty^{\mathrm r}(s)$ 
on graphs with equal rates
\begin{align}
    \tau^{\mathrm r}_\infty(s)=
    \frac{S_N-1}{b_s}\, ,
    \label{eq:taur_infinity}
\end{align}
where 
we have defined the b-degree
\begin{align}
    b_s=\sum_{s'\in {\cal B}_s} \;\sum_{k=1}^{k_{ss'}}\;
    b_{ss';k}\, .
    \label{eq:ds}
\end{align}
The b-degree $b_s$ differs from the standard
graph theoretic definition of the degree $d_s$ of state $s$.
The degree $d_s$ is defined on a graph where the 
vertices are the states and the edges are the 
moves~\cite{Boccardo2022}. Then the degree $d_s$  is the number of edges incident to
a given vertex $s$, which corresponds to the total number of
single-particle moves
from state $s$~\cite{Boccardo2022}~\footnote{
Since each move has a direction from 
state $s$ to state $s'$, our graph is a directed graph and $d_s$ is the out-degree of $s$.
For simplicity, we call $d_s$ the degree of $s$.}:
\begin{align}
    d_s=k_{ss}+\sum_{s'\in \;{\cal B}_s}k_{ss'}\, .
    \label{eq:bs}
\end{align}
Note that $k_{ss'}\geq 1$ for $s'\in {\cal B}_s$ by definition.
However, we can have $k_{ss}=0$ when $s'=s$. 
There are two differences between \cref{eq:bs,eq:ds}.
The first difference is that $b_s$
is a weighted sum with weight $b_{ss';k}$ for each move.
The second difference is that the sum over ${\cal B}_s$ discards moves that 
take the system back to the starting state $s$ in \cref{eq:bs}.

\begin{figure}
    \centering
    \includegraphics[width=\linewidth]{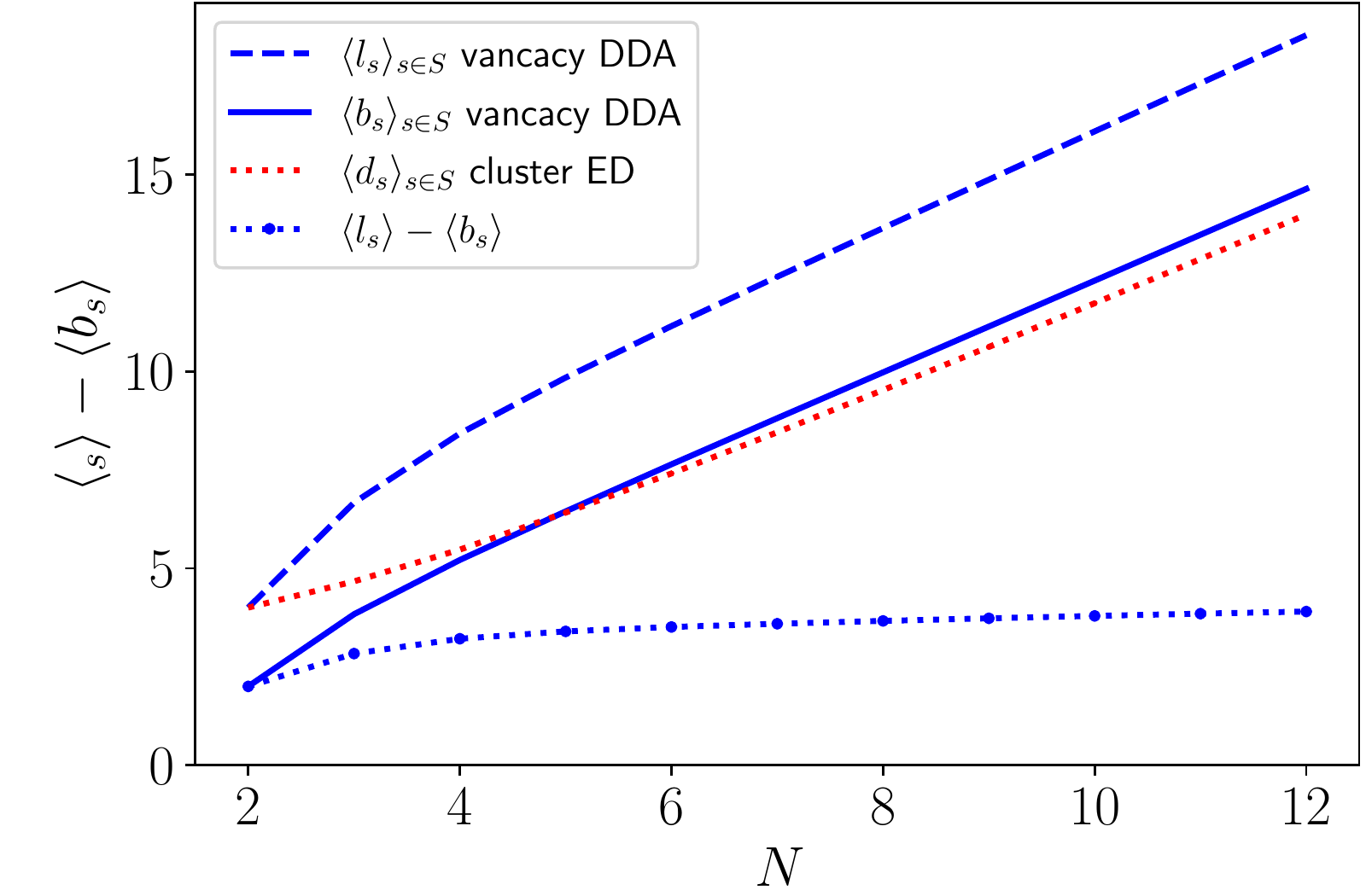}
    \caption{Average b-degree as a function of $N$. For cluster ED, we have
    $\langle d_{s} \rangle_{s\in{\cal S}}=\langle b_{s} \rangle_{s\in{\cal S}}$.
    For vacancy DDA, we show both  $\langle b_{s} \rangle_{s\in{\cal S}}$ 
    and the average number of particles $\langle \ell_s\rangle_{s\in{\cal S}}$ that can be detached
    in state $s$ as a function of $N$. 
    }
    \label{fig:b_ss}
\end{figure}

In the case of cluster ED  we have $b_{ss';k}=1$ and
there is no move which takes the system directly back to the
initial state $s$ (i.e., $k_{ss}=0$).  Hence, $b_s$ is equal to
the degree $d_s$.
Then \cref{eq:taur_infinity} reduces to the 
well known formula~\cite{Lovasz1993,Boccardo2022} $\tau^{\mathrm r}_\infty(s)=(S_N-1)/d_s$.
The average $\langle d_s\rangle_{s\in {\cal S}}$ for clusters is shown in \cref{fig:b_ss}
as a function of $N$. 

In the case of vacancy DDA, the values of $b_s$ and $d_s$ are different.
Some intuition on the physical meaning of $b_s$ can be gained by combining \cref{eq:b_ssk_DDA}
and \cref{eq:bs},
leading to~\footnote{Since there is always a pair of indices $s',k$
such that $\ell_{s;q}^\dagger=\ell_{ss';k}^\dagger$, we see that 
\cref{eq:bs_DDA} is simply a different way to sum over the moves
as compared to \cref{eq:bs}. Indeed, in the transitions
from $s$ to $s'\neq s$, we have to sum over all re-attachment
moves except those that take the system back to $s$, and we have
\begin{align}
    b_s=\sum_{s'\in {\cal B}_s} \sum_{k=1}^{k_{ss'}}
    b_{ss';k}=\sum_{\mathrm q=1}^{\ell_s} \frac{\ell_{s;q}^\dagger-1}{\ell_{s;q}^\dagger}
    \label{eq:bs_q}
\end{align}
which leads to \cref{eq:bs_DDA}.}
\begin{align}
    b_s=\ell_s-\sum_{\mathrm q=1}^{\ell_s} \frac{1}{\ell_{s;q}^\dagger}\, ,
    \label{eq:bs_DDA}
\end{align}
where $\ell_s$ is the number of particles that can be detached
in state $s$, and $q=1,..,\ell_s$ is an index for these particles.
We have also defined the number $\ell_{s;q}^\dagger$
of possible attachment sites in the excited state obtained
by removing the $q$-th particle.
We expect $\ell_{s}^\dagger$ and $\ell_{s;q}^\dagger$
to be of the same order of magnitude because they both
increase with edge length. However,
many attachment sites are forbidden in very ramified vacancies
because they would lead to breaking, so that on average
$\ell_{s;q}^\dagger$ is smaller than $\ell_{s}^\dagger$.
We therefore expect that the second term
in the right hand side of \cref{eq:bs_DDA} to be larger than $1$.
As seen in~\cref{fig:b_ss}, the difference
$\ell_s-b_s$ grows slowly 
as $N$ increases, and reaches maximum values around $4$ in the range $N\leq12$ than we have investigated.

Using \cref{eq:taur_infinity},
the return time to target is rewritten as
\begin{align}
    \tau^{\mathrm r}(s)=
    \tau^{\mathrm r}_\infty(s)
    \frac{\langle{\rm e}^{-H_{s'}/T}\rangle_{s'\in {\cal S}\backslash s}}
    {\langle\!\langle {\rm e}^{-H_{ss';k}^\dagger/T}
    \rangle\!\rangle_{k,s'\in {\cal B}_s}}
\label{eq:tau_r_Hs_Hdag_brakets}
\end{align}
where we have defined a new averaging notation for any quantity $f_{ss';k}$
\begin{align}
    {\left\langle\!\left\langle f_{ss';k}
    \right\rangle\!\right\rangle_{k,s'\in {\cal B}_s}}
= \frac{1}{b_s}\sum_{s'\in {\cal B}_s} \sum_{k=1}^{k_{ss'}}
    b_{ss';k}\,f_{ss';k}\, ,
\end{align}
that corresponds to a weighted average over
all possible moves from state $s$ with weights $b_{ss';k}$.

\begin{figure}
    \centering
    \includegraphics[width=\linewidth]{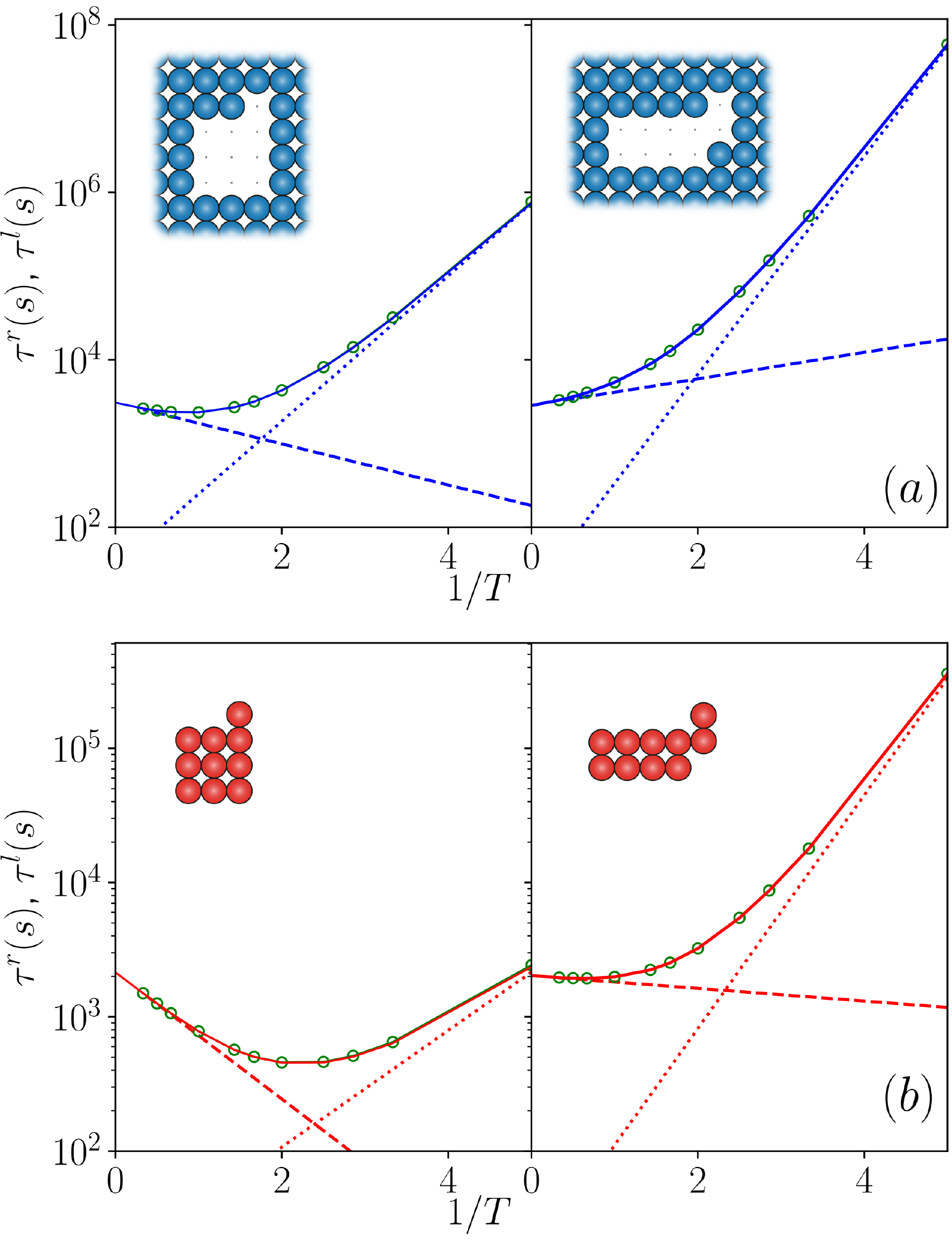}
    \caption{Expected return time $\tau^{\mathrm r}(s)$ for two clusters with $N=10$
    as a function of the inverse temperature $1/T$.
    (a) vacancy DDA. (b) cluster ED. 
    The dashed and dotted lines correspond respectively to high and low temperature expansions.
    The green empty symbols correspond to the loop time $\tau^{\ell}(s)$.
    On the left panels $H_s=7$ and $i=0$. On the right panels $H_s=8$ and $i=1$.
    }
    \label{fig:t_expansion}
\end{figure}

In \cref{fig:t_expansion}, the return
time evaluated from the equivalent expressions \cref{eq:tau_r_Hs_Hdag,eq:tau_r_Hs_Hdag_brakets} 
is shown for different cluster shapes and 
for the two models.
The expressions \cref{eq:tau_r_Hs_Hdag,eq:tau_r_Hs_Hdag_brakets} 
are in quantitative agreement with
numerical estimates based on the iterative evaluation
method reported in Ref.~\cite{Boccardo2022} for ED.
Here, we have implemented this iterative method both for
cluster ED and vacancy DDA.
Perfect agreement is found between iterative evaluation  and
the expressions based on Kac's formula \cref{eq:tau_r_Hs_Hdag,eq:tau_r_Hs_Hdag_brakets}.
Technical details on the implementation of iterative evaluation
and a comparison between the two methods 
are reported in \cref{a:recursion}.

In Ref.~\cite{Boccardo2022}, we have noticed that
$\tau^{\mathrm r}(s)$ can exhibit a minimum
as a function of the temperature $T$ for particle cluster ED. 
In \cref{fig:t_expansion}(a), we 
see that this minimum can also be present for vacancy DDA.

\subsection{High temperature expansion}
\label{s:HT_expansion}

A high temperature expansion of \cref{eq:tau_r_Hs_Hdag_brakets} to linear order in $1/T$
leads to 
\begin{align}
\tau^{\mathrm r}(s) &
\xrightarrow[T\rightarrow \infty]{}
\tau_\infty^{\mathrm r}(s) \left(1+ \frac{M^{\mathrm r}(s)}{T}\right) \, ,
\nonumber \\
M^{\mathrm r}(s)
&=
\langle\!\langle H_{ss';k}^\dagger\rangle\!\rangle_{k,s'\in {\cal B}_{s}}
-\langle H_{s'}\rangle_{s'\in {\cal S}\backslash s}
\nonumber \\
&=
\langle\!\langle H_{ss';k}^\dagger\rangle\!\rangle_{k,s'\in {\cal B}_{s}}
-\langle H_{s'}\rangle_{s'\in {\cal S}}
\nonumber \\&
+\frac{1}{S_N-1}(H_{s} - \langle H_{s'}\rangle_{s'\in {\cal S}})\, .
\label{eq:M0_of_H}
\end{align}
In the case of cluster ED, this expression is in agreement with 
the result of Ref.~\cite{Boccardo2022}.
However, this result was not written as a function of energies in Ref.~\cite{Boccardo2022}.
The derivation of \cref{eq:M0_of_H} using the method presented in Ref.~\cite{Boccardo2022} is
discussed in detailed in \cref{a:HT_M0}.

One of the main statements of
Ref.~\cite{Boccardo2022} was that the expected return time for large and compact states
exhibits a minimum at a finite temperature. Such a minimum is associated to the 
condition of a negative slope at high temperature ($1/T\rightarrow 0$) in the plots of \cref{fig:t_expansion},
i.e., $M^{\mathrm r}(s)<0$.  As $N$ increases,
 $S_N$ grows exponentially and
the terms in the last line of the expression of $M^{\mathrm r}(s)$ in \cref{eq:M0_of_H}
should be negligible. In addition, the contribution of the state $s$
to the average $\langle H_{s'}\rangle_{s'\in {\cal S}\backslash s}$
should be negligible, so that 
$\langle H_{s'}\rangle_{s'\in {\cal S}\backslash s}\approx\langle H_{s'}\rangle_{s'\in {\cal S}}$. 
Hence, the criterion $M^{\mathrm r}(s)<0$ for 
the presence of a minimum can be written approximately as
\begin{align}
    \langle\!\langle H_{ss';k}^\dagger\rangle\!\rangle_{k,s'\in {\cal B}_{s}}
<\langle H_{s'}\rangle_{s'\in {\cal S}}.
\label{ineq:optim_tau^r}
\end{align}
This condition is one of our main results.

\begin{figure}
    \centering
    \includegraphics[width=\linewidth]{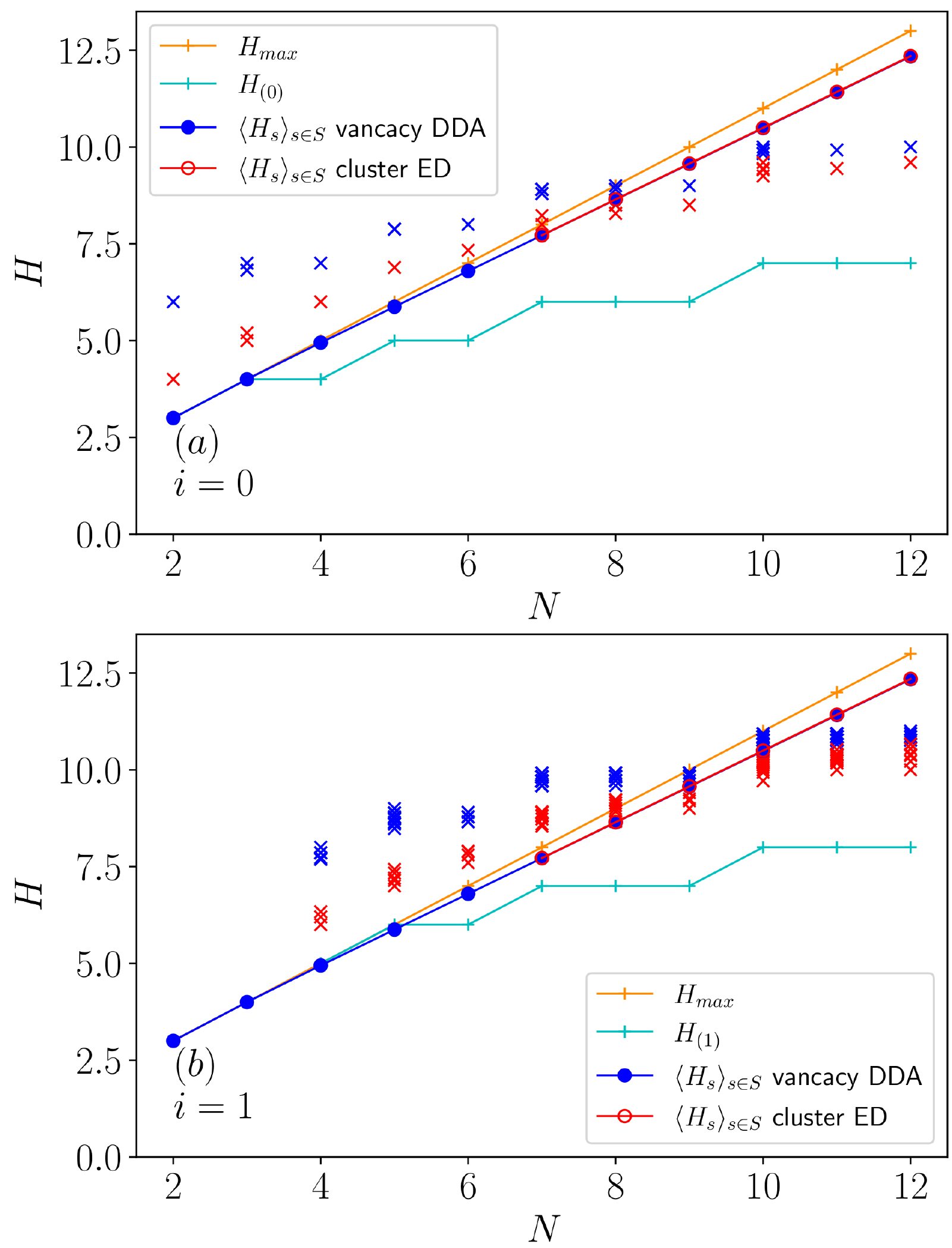}
    \caption{Comparison of different energies as a function
    of $N$ (a) for ground states with $i=0$, and
    (b) for states in the first energy level with $i=1$.
    Cyan pluses: energy $H_{(i)}$ of the states.
    Blue and red circles: average energy $\langle H_{s'}\rangle_{s'\in {\cal S}}$ 
    for vacancies and clusters respectively (these data points are almost identical). 
    Orange pluses: maximum energy $H_{max}=N+1$.
    Blue and red crosses: all possible values of $\langle\!\langle H_{ss';k}^\dagger\rangle\!\rangle_{k,s'\in {\cal B}_{s}}$.
    }
    \label{fig:E_mean}
\end{figure}

In \cref{fig:E_mean}(a,b), we have reported the average energy $\langle H_{s'}\rangle_{s'\in {\cal S}}$
as a function of $N$. Since higher energy levels have higher degeneracy $G_{(i)}$
(as seen from \cref{fig:degeneracy_Gi}), the value of $\langle H_{s'}\rangle_{s'\in {\cal S}}$ is close to
the maximum possible value of the energy $H_{max}=H_{(i_{\mathrm max})}=N+1$.

In addition, the values of $\langle\!\langle H_{ss';k}^\dagger\rangle\!\rangle_{k,s'\in {\cal B}_{s}}$
for all possible transitions to states $s'$  are shown on \cref{fig:E_mean}(a)
when starting from a ground state $s$ with $i=0$,
and on \cref{fig:E_mean}(b) when starting from the first excited state $s$ with $i=1$.
As a first remark, due to the average curvature
of the interface, atoms that detach from the edge of vacancies have on average more
bonds to break. Thus the excited state energies of vacancies are higher. 
This is seen for $i=0$ and $i=1$ in \cref{fig:E_mean}.

A close inspection of \cref{fig:E_mean}(a,b) shows that for a given $i$, 
a given $N$ and a given model (cluster ED or vacancy DDA),
the values of 
$\langle\!\langle H_{ss';k}^\dagger\rangle\!\rangle_{k,s'\in {\cal B}_{s}}$
are either all above $\langle H_{s'}\rangle_{s'\in {\cal S}}$, 
or all below $\langle H_{s'}\rangle_{s'\in {\cal S}}$.
Moreover, the value of $N$ above which $\langle\!\langle H_{ss';k}^\dagger\rangle\!\rangle_{k,s'\in {\cal B}_{s}}$
is below $\langle H_{s'}\rangle_{s'\in {\cal S}}$ depends on $i$ and on the model.
For cluster ED, a minimum is predicted when $N\geq 8$ with $i=0$, and when $N\geq 9$ with $i=1$.
For vacancy DDA, a minimum is predicted when $N\geq 9$ with $i=0$, and when $N\geq 11$ with $i=1$.

These conditions for $i=0$ and $i=1$ are in agreement with the observation
of a minimum in the return time of the ground states.
For example let us consider \cref{fig:t_expansion},
where $N=10$, $i=0$ on the left panels and $i=1$ on the right panels.
A minimum is observed  for clusters with $i=0$ and $i=1$,
and for vacancies only when $i=0$,
in agreement with the above statements.

\subsection{Low temperature expansion}

The expression \cref{eq:tau_r_Hs_Hdag} also allows one
to obtain an asymptotic low-temperature expression
for the expected return time.
For states $s$ that are not a ground state, i.e., $H_s>H_{(0)}$, we have
\begin{align}
    \tau^{\mathrm r}(s)\xrightarrow[T\rightarrow 0]{}
    \frac{G_{(0)}}{G_{s{(0)}}^\dagger}{\rm e}^{(H_{s{(0)}}^\dagger-H_{(0)})/T} \, ,
\label{eq:tau_r_Hs_Hdag_LT}
\end{align}
where we recall that $G_{(0)}$ is the number of different configurations
that correspond to the ground state with energy $H_{(0)}$.
Moreover,  we have defined $H_{s{(0)}}^\dagger$ as the lowest energy over all excited
states of a state $s$, and
\begin{align}
    G_{s{(0)}}^\dagger=\sum_{s'\in {\cal B}_s}\;\; \sum_{k\,|\,H_{ss';k}^\dagger=H_{s{(0)}}^\dagger}
    b_{ss';k}\, ,
\end{align}
where the sum on $k$ is performed only on the moves
that take the cluster to the excited state with the
lowest energy, i.e., $H_{ss';k}^\dagger=H_{s{(0)}}^\dagger$.

When the state $s$ is a ground state, i.e., $H_s=H_{(0)}$ the
contribution that dominates the denominator of \cref{eq:tau_r_Hs_Hdag} at low temperature
depends on the fact that the ground state is unique or not.
If the ground state is not unique, i.e. if $G_{(0)}\geq 2$,
then
\begin{align}
    \tau^{\mathrm r}(s)\xrightarrow[T\rightarrow 0]{}
    \frac{G_{(0)}-1}{G_{s{(0)}}^\dagger}{\rm e}^{(H_{s{(0)}}^\dagger-H_{(0)})/T} .
\label{eq:tau_r_Hs_Hdag_LT_non_unique_GroundState}
\end{align}
However, if the ground-state is unique, i.e. $G_{(0)}=1$, the
contribution that dominates the denominator of \cref{eq:tau_r_Hs_Hdag} at low temperatures
is the energy level $i=1$ just above the ground state.
Using \cref{eq:energy_levels}, we then find
\begin{align}
    \tau^{\mathrm r}(s)\xrightarrow[T\rightarrow 0]{}
    \frac{G_{(1)}}{G_{s{(0)}}^\dagger}{\rm e}^{(H_{s{(0)}}^\dagger-H_{(0)}-1)/T}\, .
\label{eq:tau_r_Hs_Hdag_LT(0)}
\end{align}
Since this latter case of a unique ground-state only occurs for
square clusters, \cref{eq:tau_r_Hs_Hdag_LT(0)} only applies 
in the specific case of square clusters.

The low-temperature expansion \cref{eq:tau_r_Hs_Hdag_LT,eq:tau_r_Hs_Hdag_LT_non_unique_GroundState},
are shown in \cref{fig:t_expansion} for a state in the energy level $i=1$ (right panel) and
for a non-unique ground-state (left panels).

\subsection{Return time optimal temperature}

\begin{figure}
    \centering
    \includegraphics[width=\linewidth]{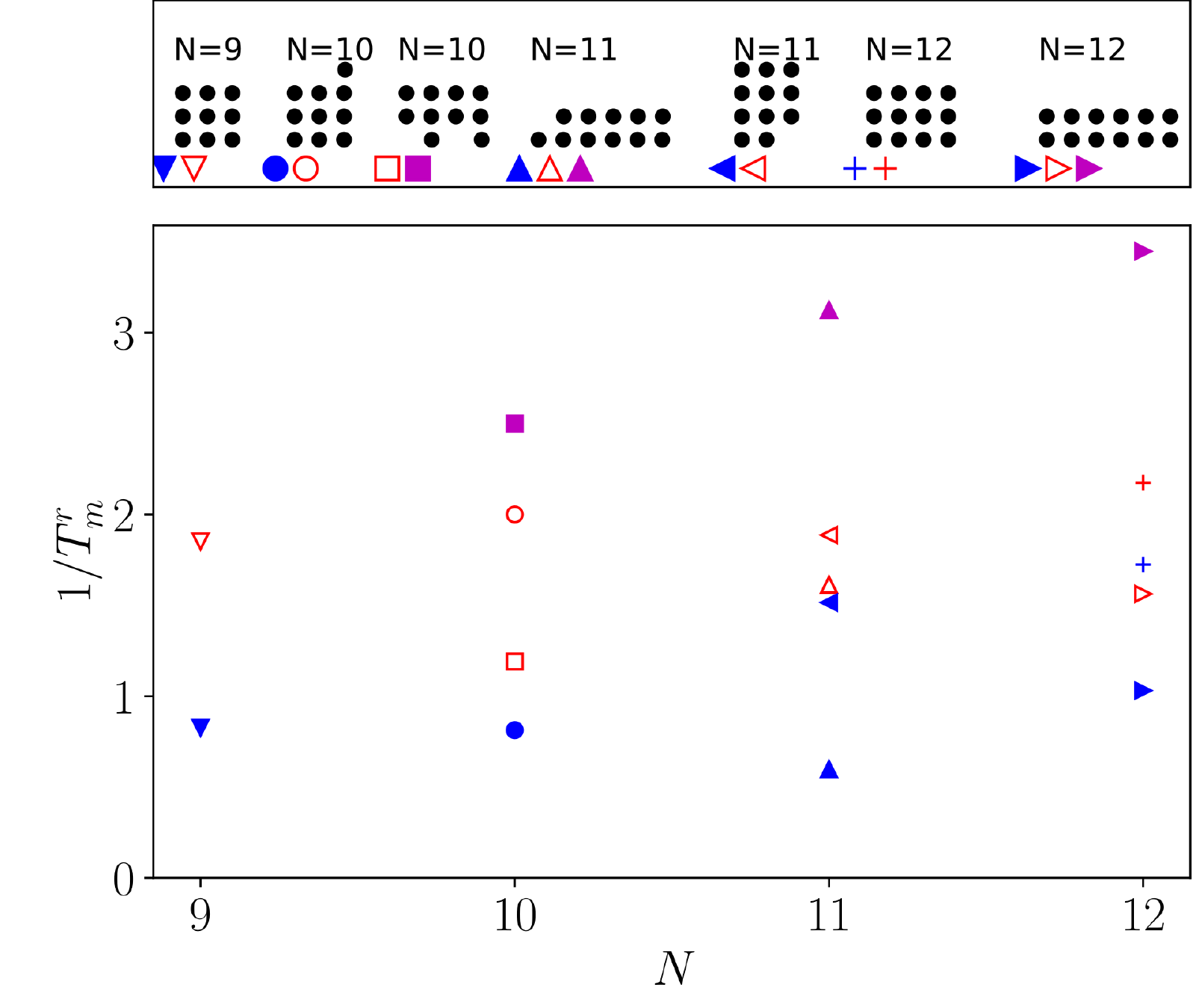}
    \caption{Inverse optimal return time temperature $1/T_m^r(s)$ for the states
    listed in the top panel.
    Blue (dark) symbols: vacancy DDA.
    Red (empty) symbols: cluster ED. 
    Violet (lighter) symbols: inverse of the equilibrium optimal temperature $1/T_m^{eq}$ when it is finite
    ($T_m^{eq}$ exhibits very similar values for vacancy DDA and cluster ED).
    For ground states ($i=0$), corresponding to the 1st, 2nd, 5th and 6th clusters
    on the top panel, $1/T_m^{eq}\rightarrow\infty$ is not shown.
    }
    \label{fig:t_optimal_tau}
\end{figure}

The optimal return time temperature 
is the temperature at which $\tau^{\mathrm r}(s)$ is minimum.
This temperature is a kinetic quantity that depends 
on the dynamical properties of the system.
When $\tau^{\mathrm r}(s)$ exhibit an optimal temperature $T_{m}^{\mathrm r}(s)$
where it is minimum, then $\partial_T\tau^{\mathrm r}(s)=0$. 
Applying this condition on \cref{eq:tau_r_Hs_Hdag_brakets}
leads to
\begin{align}
    &\frac{\left\langle H_{s'}{\rm e}^{-H_{s'}/T_m^{\mathrm r}(s)}\right\rangle_{s'\in {\cal S}\backslash s}}
    {\left\langle{\rm e}^{-H_{s'}/T_m^{\mathrm r}(s)}\right\rangle_{s'\in {\cal S}\backslash s}}
    =\left.\frac{\overline{H_{eq}}-H_sP_{eq}(s)}
    {1-P_{eq}(s)}\right|_{T=T_m^{\mathrm \ell}(s)}
\nonumber \\
    &=
    \frac
    {\left\langle\!\!\left\langle H_{ss';k}^\dagger\,{\rm e}^{-H_{ss';k}^\dagger/T_m^{\mathrm r}(s)}
    \right\rangle\!\!\right\rangle_{k,s'\in {\cal B}_s}}
    {\left\langle\!\!\left\langle {\rm e}^{-H_{ss';k}^\dagger/T_m^{\mathrm r}(s)}
    \right\rangle\!\!\right\rangle_{k,s'\in {\cal B}_s}}\, ,
    \label{eq:optimal_temp}
\end{align}
where the expression in the right hand side of the first line
is a direct rewriting of the left hand side.
In \cref{fig:t_optimal_tau}, the optimal return time inverse temperature $1/T_m^{\mathrm r}(s)$
is plotted for various states $s$ with different size $N$.
These temperatures are obtained by means of a
direct numerical estimate of the temperature at which $\tau^{\mathrm r}(s)$
is minimum.

We observe that the values of $T_m^{\mathrm r}(s)$ in \cref{fig:t_optimal_tau}
are actually higher than the corresponding equilibrium optimal temperatures $T_m^{eq}(s)$.
This can be explained from an inspection of \cref{eq:return_res_eq}
which states that $\tau^{\mathrm r}(s)$ is the product of $t(s)$
with $1/P_{eq}(s)-1$. Indeed,  $T_m^{\mathrm eq}(s)$
corresponds to a minimum of the factor $1/P_{eq}(s)-1$. Since $P_{eq}(s)$ is very similar
for particle-cluster and vacancy-clusters as discussed in \cref{s:equil_distrib}, this factor
can be considered to be identical in these two cases for our qualitative discussion. 
In addition, all rates $\gamma(s,s')$ decrease with temperature
due to their Arrhenius form, so that $t(s)$ which is the inverse of a sum of rates
from \cref{eq:t0_gamma}, increases with decreasing temperature. 
In general, the product of a monotonic function with a function with a minimum
leads to a function with a shifted minimum.
Here, due to the monotonic decrease of the residence time $t(s)$ with temperature, 
the minimum $T_m^{\mathrm r}(s)$ of the product $t(s)(1/P_{eq}(s)-1)$ is shifted
to a higher temperature as compared to the minimum $T_m^{\mathrm eq}(s)$ of $(1/P_{eq}(s)-1)$. 
As a consequence,  $T_m^{\mathrm r}(s)$
is always higher than $T_m^{eq}(s)$:
\begin{align}
    T_m^{\mathrm r}(s)>T_m^{eq}(s).
    \label{ineq:shift_optimal}
\end{align}
This inequality is confirmed by \cref{fig:t_optimal_tau},
where the inverse equilibrium temperature $1/T_m^{eq}(s)$
is either infinite for ground states (as discussed in \cref{s:equil_optimal_temperature},
a ground state can be considered as a state with $T_m^{eq}(s)=0$), or finite and plotted in purple
for non-ground states.
We see in \cref{fig:t_optimal_tau} that the associated 
inverses of the return time optimal temperatures $1/T_m^{\mathrm r}(s)$
are always smaller than $1/T_m^{eq}(s)$.

Because of this shift toward higher temperatures,
low temperature approximations for $T_m^{\mathrm r}(s)$ are not accurate.
Nevertheless, a high-temperature expansion
can be performed from \cref{eq:optimal_temp}.
Its detailed expression
and a comparison with the true optimal
temperature is reported in \cref{a:HT_optim}.

Finally, another general feature of $T_m^{\mathrm r}(s)$
can be observed.
Indeed, as already noticed in \cref{s:HT_expansion},
the excited state energies of vacancies are higher due to the curvature of the edge. 
As a consequence of their higher excited state energies,  
the decrease of $t(s)$  with temperature (see \cref{eq:t0_gamma,eq:rates_Hdag_Hs})
is actually faster for vacancies.
Hence, the shift of the optimal temperature
towards higher temperatures is stronger for 
vacancies, leading to a higher return time optimal
temperature $T_m^{\mathrm r}(s)$.
This effect is systematically observed in \cref{fig:t_optimal_tau}.

\subsection{Comparison between $\tau^{\mathrm{r}}(s)$ and $\tau^{\mathrm{\ell}}(s)$}

The expected loop time exhibits an expression similar to
that of the return time
\begin{align}
    \tau^{\mathrm{\ell}}(s)=
    \tau^{\mathrm{\ell}}_\infty(s)
    \frac{\langle{\rm e}^{-H_{s'}/T}\rangle_{s'\in {\cal S}}}
    {\langle\!\langle {\rm e}^{-H_{ss';k}^\dagger/T}
    \rangle\!\rangle_{k,s'\in {\cal B}_s}},
\label{eq:tau_r_Hs_Hdag_brakets_loop}
\end{align}
where the infinite temperature expected loop time is
\begin{align}
    \tau^{\mathrm {\ell}}_\infty(s)=
    \frac{S_N}{b_s}\, .
    \label{eq:tauell_infinity}
\end{align}
Moreover, the high temperature expansion 
\begin{align}
\tau^{\mathrm {\ell}}(s) &
\xrightarrow[T\rightarrow \infty]{}\tau_\infty^{\mathrm {\ell}}(s) \left(1+ \frac{M^{\ell}(s)}{T}\right) \, ,
\nonumber \\
M^{\ell}(s)
&=
\langle\!\langle H_{ss';k}^\dagger\rangle\!\rangle_{k,s'\in {\cal B}_{s}}
-\langle H_{s'}\rangle_{s'\in {\cal S}}.
\label{eq:M0ell_of_H}
\end{align}
is very similar to that of the return time.
The two expansions \cref{eq:M0_of_H,eq:M0ell_of_H} are actually identical to leading order in $1/S_N$.
Thus, the condition for the presence of an optimal temperature \cref{ineq:optim_tau^r}
is also valid for the loop time (it is actually exact for the loop time,
while it was only an approximation for the return time).

At low temperatures, the asymptotic behaviors
of $\tau^{\mathrm {\ell}}(s)$ and $\tau^{\mathrm r}(s)$
are also identical, except when $s$ is the ground state.
For all states including the ground states, we have at low temperatures
\begin{align}
    \tau^{\mathrm {\ell}}(s)\xrightarrow[T\rightarrow 0]{}
    \frac{G_{(0)}}{G_{s{(0)}}^\dagger}{\rm e}^{(H_{s{(0)}}^\dagger-H_{(0)})/T} \, .
\label{eq:tau_ell_Hs_Hdag_LT}
\end{align}

Furthermore, there is also an optimal temperature 
at which the loop time can be minimum. The equation
for $T_m^\ell(s)$ is similar to \cref{eq:optimal_temp} and reads
\begin{align}
    \left.\overline{H_{eq}}\right|_{T=T_m^{\mathrm \ell}(s)}=
    \frac
    {\left\langle\!\!\left\langle H_{ss';k}^\dagger\,{\rm e}^{-H_{ss';k}^\dagger/T_m^{\mathrm \ell}(s)}
    \right\rangle\!\!\right\rangle_{k,s'\in {\cal B}_s}}
    {\left\langle\!\!\left\langle {\rm e}^{-H_{ss';k}^\dagger/T_m^{\mathrm \ell}(s)}
    \right\rangle\!\!\right\rangle_{k,s'\in {\cal B}_s}}\, .
    \label{eq:optimal_temp_eq_loop}
\end{align}
Globally, $\tau^{\mathrm {\ell}}(s)$ and $\tau^{\mathrm r}(s)$
are expected to be similar when $N$ is large enough,
and when $s$ is not a unique a ground state.

\begin{figure}
    \centering
    \includegraphics[width=\linewidth]{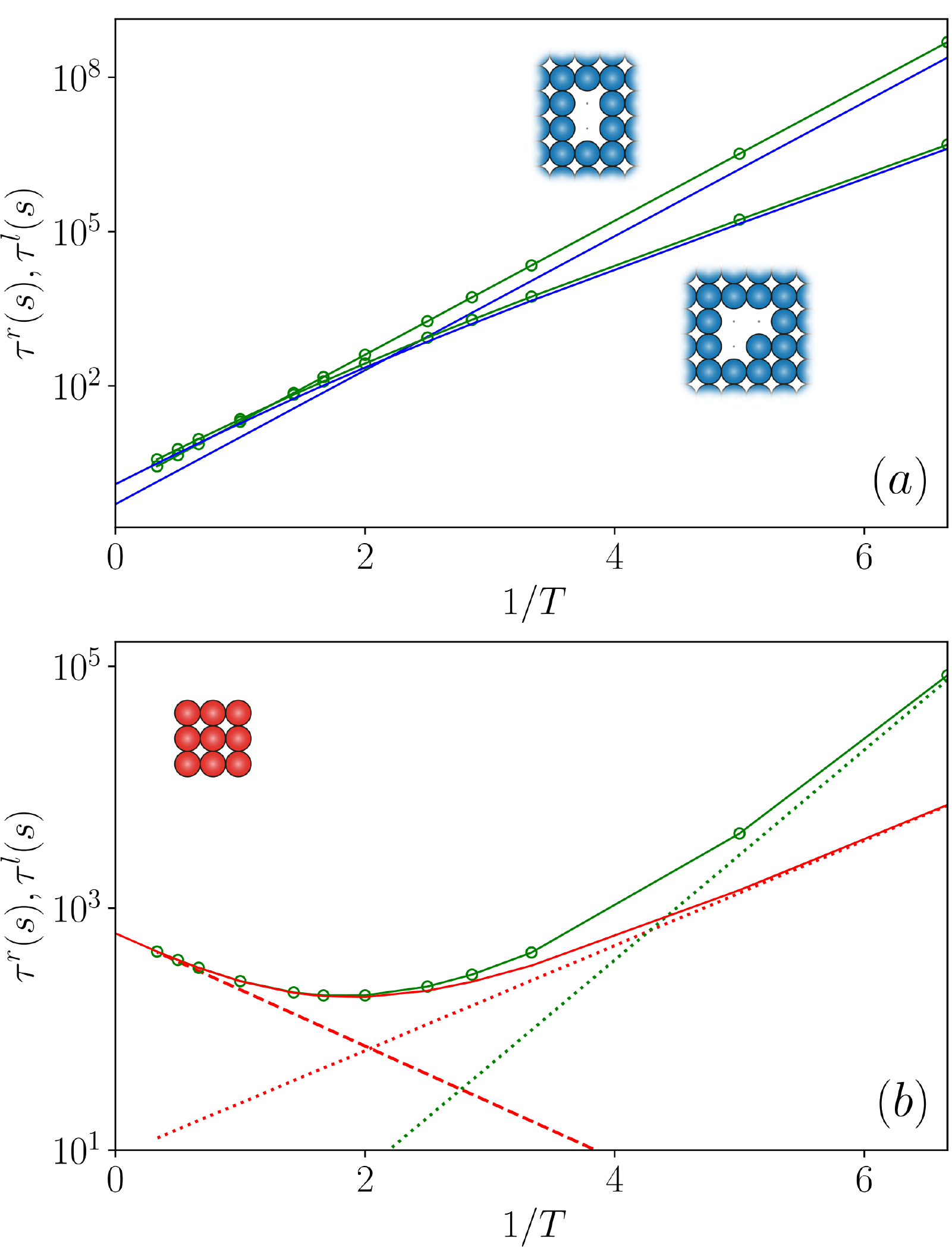}
    \caption{Differences between expected return time $\tau^{\mathrm {\ell}}(s)$ and loop time $\tau^{\mathrm r}(s)$.
    (a) Small vacancies with $N=2$ and $N=3$, vacancy DDA.
    (b) Square cluster with $N=9$, cluster ED.
    }
    \label{fig:t_expansion_ED}
\end{figure}

In \cref{fig:t_expansion_ED}, we focus on cases
where a significant difference can be found 
between $\tau^{\mathrm {\ell}}(s)$ and $\tau^{\mathrm r}(s)$.
The cases of dimers $N=2$ and trimers $N=3$
are reported in \cref{fig:t_expansion_ED}(a).
In these special cases with a single energy level (i.e., $i_{max}=0$)
we have a simple relation independent of the state and of the temperature $\tau^\ell(s)/\tau^r(s)=1/(1-1/S_N)$, which leads
to $\tau^\ell(s)/\tau^r(s)=2$ for $N=2$ and $\tau^\ell(s)/\tau^r(s)=6/5$ for $N=3$.
Another case where $\tau^{\mathrm {\ell}}(s)$ and $\tau^{\mathrm r}(s)$
can be different is that of square islands at low temperature,
as shown in \cref{fig:t_expansion_ED}(b).

\section{Discussion and Conclusion}

To conclude, we have reported on the properties of
the expected return time $\tau^{\mathrm r}(s)$ to a given cluster
configuration $s$ in equilibrium.
We have focused on a broken-bond model
that allows one to describe edge-diffusion (ED) for clusters and
detachment-diffusion-re-attachment (DDA) inside vacancies
within a unified framework.
The evaluation of $\tau^{\mathrm r}(s)$ from the residence time
and the equilibrium probability distribution 
is much faster numerically than the evaluation 
based on the iterative evaluation
method presented in Ref.~\cite{Boccardo2022}.
In addition, the analysis of this expression
leads to simple and intuitive expansions
in the high and low temperature regimes, 
and allows one to study the optimal temperature at which $\tau^{\mathrm r}(s)$ is minimum.
The origin of this optimal temperature can be traced back to the
equilibrium optimal temperature at which the probability of observing
a given low-energy state is maximum. The return time optimal temperature is
found to be shifted to higher temperatures as compared to the equilibrium
optimal temperature. This shift is larger for vacancies than for islands.

We hope that the investigation of the properties of cluster return times
will provide useful hints for a better understanding of first passage times.
Indeed, their properties are similar to those of first passage
times but are much simpler to analyze.
We hope that our work will motivate theoreticians
and experimentalists to investigate 
cluster return times with various types of mass transport kinetics.

\begin{appendix}

\section{Recursion relation for fist passage times
and iterative evaluation}
\label{a:recursion}

The expected first passage time $\tau(s,{\bar{s}})$
from state $s$ to state $\bar{s}$ obeys a recursion relation~\cite{Boccardo2022}
\begin{align}
\label{eq:recursion}
\tau(s,{\bar{s}}) = t(s) +\! \sum_{s'\in {\cal B}_s}\!p(s,s')\tau(s',{\bar{s}})\,, 
\end{align}
where 
\begin{align}
\label{aeq:transition_proba}
p(s,s') = t(s) \gamma(s,s')\,. 
\end{align}
In Ref.\cite{Boccardo2022}, we have reported on the study
of first passage times from an arbitrary state $s$ to another 
arbitrary state $\bar{s}$, called the target state.

The expected return time $\tau^{\mathrm r}(\bar{s})$  to the state $\bar{s}$
can be written as
a sum of the first passage times over the neighbors of $\bar{s}$ \cite{Boccardo2022}
\begin{align}
\label{aeq:3}
\tau^{\mathrm r}(\bar{s}) 
&= \sum_{s\in {\cal B}_{\bar{s}}}p(\bar{s},s)\tau(s,\bar{s})\, .
\end{align}

\begin{figure}
    \centering
    \includegraphics[width=\linewidth]{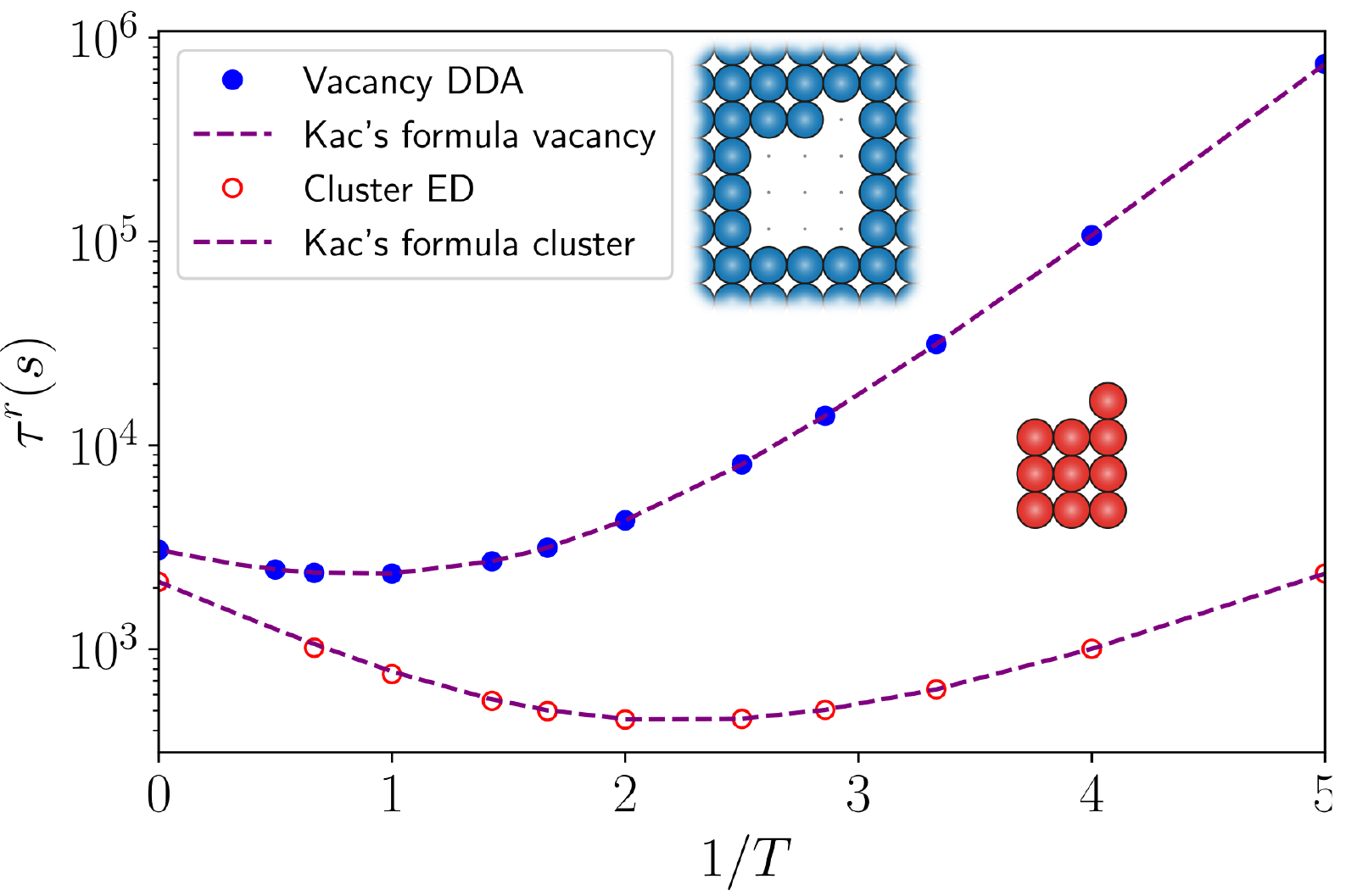}
    \caption{Comparison of the return time $\tau^r(s)$
    from Kac's formula \cref{eq:return_res_eq} and the 
    value obtained by iterative evaluation.
    }
    \label{fig:dp_value}
\end{figure}

As discussed in Ref.~\cite{Boccardo2022}, a numerical solution of $\tau(s,{\bar{s}})$ can be obtained 
via a simple iterative evaluation of \cref{eq:recursion}.
Then, $\tau^{\mathrm r}(\bar{s})$ is obtained from \cref{aeq:3}.
In \cref{fig:dp_value}, this iterative method is shown to be in quantitative agreement
with Kac's formula \cref{eq:return_res_eq} for the cluster ED and vacancy DDA models.
However, the iterative method is slower because it
requires to determine $\tau(s,{\bar{s}})$ for all states $s$.

\section{High temperature expansion
link to the results of Ref.\cite{Boccardo2022}}
\label{a:HT_M0}

\subsection{High temperature expansion}

In Ref.\cite{Boccardo2022},
we have derived an expression that is convenient for the study of the high temperature regime
\begin{align}
\label{aeq:4}
\tau^{\mathrm r}(&\bar{s}) 
=t(\bar{s})(S_N-1)
\nonumber \\
&+t(\bar{s})\sum_{s}\tau(s,\bar{s})
\sum_{s'\in {\cal B}_s}(\gamma(s',s)-\gamma(s,s'))\, .
\end{align}

Following the same lines as in  Ref.\cite{Boccardo2022},
we  perform an expansion to linear order in $1/T$ to obtain the first correction to Eq.~(\ref{eq:taur_infinity}).
We start with the expansion of the rates
\begin{align}
\label{eq:expansion_gamma_HT} 
    \gamma(s,s')=b_{ss'}-\frac{n_{ss'}}{T},
\end{align}
where for the sake of concision, we have defined 
\begin{align}
    b_{ss'}=b_{s's}&=\sum_{k=1}^{k_{ss'}}b_{ss';k}\, ,
    \\
\label{eq:tilde_n}
    {n}_{ss'}&=\sum_{k=1}^{k_{ss'}}b_{ss';k}\, n_{ss';k}\, .
\end{align}

Inserting this expression in the expected return time Eq.~(\ref{aeq:4}), we obtain
to linear order in $1/T$
\begin{align}
\label{aeq:7}
\tau^{\mathrm r}(\bar{s}) &=\tau_\infty^{\mathrm r}(\bar{s}) \left(1+ \frac{M^{\mathrm r}(\bar{s})}{T}\right) 
\nonumber \\
M^{\mathrm r}(\bar{s})&=\frac{1}{b_{\bar{s}}}\sum_{s\in {\cal B}_{\bar{s}}} {n}_{\bar{s}s}
\nonumber \\&
+\frac{1}{S_N-1}\sum_{s}\tau_\infty(s,\bar{s})
\sum_{s'\in {\cal B}_s}({n}_{ss'}-{n}_{s's})
\, ,
\end{align}
where $\tau_\infty(s,\bar{s})$ is the value of 
$\tau(s,\bar{s})$ when $T\rightarrow\infty$.

\subsection{Expression of $M^{\mathrm r}(s)$ as a function of energies}

Using \cref{eq:n_H}, $M^{\mathrm r}$ is rewritten as
\begin{align}
M^{\mathrm r}(\bar{s})
&=-H_{\bar{s}}+\frac{1}{b_{\bar{s}}}\sum_{s\in {\cal B}_{\bar{s}}} {H}_{\bar{s}s}^\dagger
\nonumber \\&
-\frac{1}{S_N-1}\sum_{s}
\sum_{s'\in {\cal B}_s}b_{ss'}(\tau_\infty(s,\bar{s})H_{s}-\tau_\infty(s,\bar{s})H_{s'})
\label{aeq:M_energies}
\end{align}
where we have used Eq.(\ref{aeq:symm_H_dagger}),
and we have defined
\begin{align}
\label{eq:tilde_Hdag}
    {H}_{ss'}^\dagger=\sum_{k=1}^{k_{ss'}}b_{ss';k}H_{ss';k}^\dagger\, .
\end{align}

We now notice that the double sum in the last line of \cref{aeq:M_energies} corresponds to 
a sum over all possible physical moves from $s$ to $s'$.
We can therefore exchange $s$ and $s'$ in the last term of the sum, leading to
\begin{align}
\label{aeq:M0_intermediate}
M^{\mathrm r}(\bar{s})
&=-H_{\bar{s}}+\frac{1}{b_{\bar{s}}}\sum_{s\in {\cal B}_{\bar{s}}} {H}_{\bar{s}s}^\dagger
\nonumber \\&
-\frac{1}{S_N-1}\sum_{s}H_{s}
\sum_{s'\in {\cal B}_s}b_{ss'}(\tau_\infty(s,\bar{s})-\tau_\infty(s',\bar{s})).
\end{align}

Now, we notice that the recursion relation \cref{eq:recursion}
can be re-written as
\begin{align}
    1=\sum_{s'\in {\cal B}_s}\gamma(s,s')(\tau(s,\bar{s})-\tau(s',\bar{s}))\, .
\end{align}
for any $s\neq \bar{s}$.
In the limit of infinite temperatures, this equation reads
\begin{align}
\label{eq:Laplacian_HT}
    1=\sum_{s'\in {\cal B}_s}b_{ss'}(\tau_\infty(s,\bar{s})-\tau_\infty(s',\bar{s})) \,.
\end{align}
The last term \cref{aeq:M0_intermediate} 
is seen to be equal to $1$ from Eq.(\ref{eq:Laplacian_HT})
for all $s\neq \bar{s}$, so that
\begin{align}
M^{\mathrm r}(\bar{s})
&=-H_{\bar{s}}+\frac{1}{b_{\bar{s}}}\sum_{s\in {\cal B}_{\bar{s}}} {H}_{\bar{s}s}^\dagger
\nonumber \\&
-\frac{1}{S_N-1}\sum_{s}H_{s}+\frac{1}{S_N-1}H_{\bar{s}}
\nonumber \\&
+\frac{1}{S_N-1}H_{\bar{s}}
\sum_{s'\in {\cal B}_{\bar{s}}}b_{\bar{s}s'}\tau_\infty(s',\bar{s}).
\label{aeq:M_energies_bis}
\end{align}

From \cref{eq:expansion_gamma_HT} the rate at infinite temperature is $\gamma_\infty(s,s')=b_{ss'}$,
and the residence time is $t_\infty(s)=1/(\sum_{\in {\cal B}_{s}}b_{ss'})=1/b_s$.
Thus, using  \cref{aeq:transition_proba,aeq:3,eq:taur_infinity} we have
\begin{align}
\sum_{s'\in {\cal B}_{\bar{s}}}b_{\bar{s}s'}\tau_\infty(s',\bar{s})
&=b_{\bar{s}}\sum_{s'\in {\cal B}_{\bar{s}}}p_\infty(\bar{s},s')\tau_\infty(s',\bar{s})
\nonumber \\
&=b_{\bar{s}}\tau_\infty^{\mathrm r}(\bar{s})=S_N-1.
\end{align}
Using this relation in \cref{aeq:M_energies_bis}, we find
\begin{align}
M^{\mathrm r}(\bar{s})
&=
\frac{1}{b_{\bar{s}}}\sum_{s\in {\cal B}_{\bar{s}}} {H}_{\bar{s}s}^\dagger
+\frac{1}{S_N-1}(H_{\bar{s}}-\sum_{s}H_{s})
\end{align}
Recombining the terms of this latter equation leads to \cref{eq:M0_of_H}.

\section{High temperature expansion of the optimal temperatures}
\label{a:HT_optim}

\begin{figure}
    \centering
    \includegraphics[width=\linewidth]{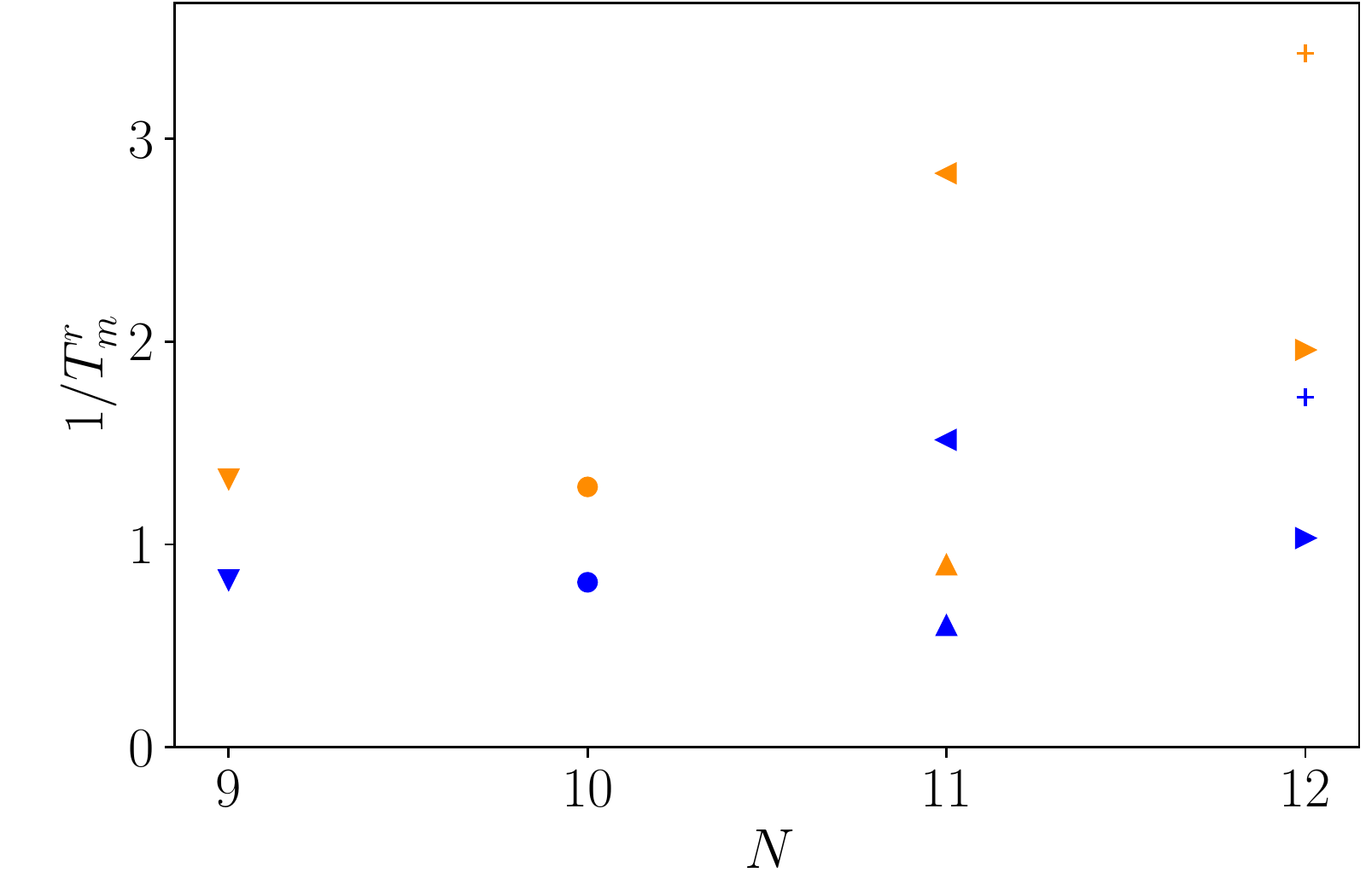}
    \caption{Optimal return time inverse temperature $1/T_m^{\mathrm r}(s)$ for various states with vacancy DDA.
    The prediction from the high temperature expansion
    \cref{eq:optimal_return_temp_HT} is shown in orange (light) symbols.
    }
    \label{fig:t_optimal_tau_2}
\end{figure}

A first-order expansion of \cref{eq:optimal_temp} in  $1/T_m^{\mathrm r}(s)$
leads to
\begin{widetext}
\begin{align}
    T_{m,HT}^{\mathrm r}(s)
    =
    \frac
    {\langle H_{s'}^2\rangle_{s\in{\cal S}\backslash s}-\langle H_{s'}\rangle^2_{s\in{\cal S}\backslash s} 
    -\langle\!\langle H_{ss';k}^{\dagger2}\rangle\!\rangle_{k,s'\in {\cal B}_s}
    +\langle\!\langle H_{ss';k}^{\dagger}\rangle\!\rangle^2_{k,s'\in {\cal B}_s}} 
    {\langle H_{s'}\rangle_{s\in{\cal S}\backslash s}
    -\langle\!\langle H_{ss';k}^\dagger\rangle\!\rangle_{k,s'\in {\cal B}_s}
    }
    \label{eq:optimal_return_temp_HT}
\end{align}
\end{widetext}
On \cref{fig:t_optimal_tau_2}, the resulting estimate of the optimal temperature 
is seen to be in reasonable agreement with the observed minimum of $\tau^{\mathrm{r}}(s)$
for vacancy DDA, and to be become better as the optimal temperature increases.
This agreement is expected because the optimal temperatures are rather high.
For the same reason, the other estimates based on a low temperature expansion
and on the matching between low and high temperature expansions are inaccurate
in the range of size $N$ that we have explored.

\end{appendix}


\bibliography{references}


\end{document}